\documentclass[a4paper,11pt]{article}
\pdfoutput=1 % if your are submitting a pdflatex (i.e. if you have
\usepackage{jcappub} % for details on the use of the package, please
\usepackage[T1]{fontenc} % if needed
\usepackage{orcidlink}
\usepackage{newtxtext,newtxmath}
\usepackage{amsmath}
\usepackage{physics}
\usepackage{cleveref}
\usepackage{comment}
\usepackage{multicol,multirow}

\newcommand{\Msun}[1]{\ensuremath{\,h^{-1}\,\textup{M}_\odot}}
\newcommand{\mpch}{\,h^{-1}\,{\rm Mpc}}
\newcommand{\invmpch}{\,h\,{\rm Mpc}^{-1}}
\newcommand{\gpch}{\,h^{-3}\,{\rm Gpc}^{3}}
\newcommand{\degsq}{\,{\rm deg}^{2}}
\newcommand{\kms}{\,{\rm km}\,{\rm s}^{-1}}

\newcommand{\As}{\text{log}(10^{10}A_s)}
\newcommand{\oii}{[O\,\textsc{ii}] }
\newcommand{\oiirel}{[\text{O}\,\textsc{ii}]_R(z) }
\newcommand{\oiii}{[O\,\textsc{iii}] }

\newcommand{\Yone}[0]{DR1 }
\newcommand{\full}[0]{\texttt{full} }
\newcommand{\abacus}[0]{\textsc{AbacusSummit} }

\newcommand{\elg}[0]{ELG }
\newcommand{\tsnr}[0]{S_{\rm spec,obs} }
\newcommand{\tsnrtheory}[0]{S_{\rm spec,theo} }
\newcommand{\tsnrmean}[0]{\overline{T}_{\rm ELG} }
\newcommand{\SSR}[0]{f_{\rm goodz} }
\newcommand{\SSRmodel}[0]{f_{\rm goodz}^{\rm m} }
\newcommand{\wzfail}[0]{w_{\rm zfail} }
\newcommand{\czfail}[0]{\eta_{\rm zfail} }
\newcommand{\DN}[0]{BASS/MzLS }
\newcommand{\DS}[0]{DECam }
\newcommand{\fluxg}[0]{\text{fluxG} }

\crefname{equation}{Eq.}{Eqs.}
\crefname{section}{Section}{Sections}
\crefname{figure}{Figure}{Figures}
\crefname{table}{Table}{Tables}
\crefname{appendix}{Appendix}{Appendices}
\Crefname{figure}{Figure}{Figures}
\Crefname{equation}{Equation}{Equations}
\Crefname{section}{Section}{Sections}
\Crefname{table}{Table}{Tables}

\defcitealias{mine}{Yu23}

\newcommand{\AGK}[1]{\textcolor{purple}{[\textbf{Alex}: #1]}}
\newcommand{\jiaxi}[1]{{\textcolor{cyan}{[YJX: #1]}}}

\title{ELG Spectroscopic Systematics Analysis of the DESI Data Release 1
}

\affiliation{Affiliations are in Appendix \ref{appendix:affiliations}}

\author[1]{{J.~Yu}\orcidlink{0009-0001-7217-8006},}
\author[2,3,4]{{A.~J.~Ross}\orcidlink{0000-0002-7522-9083},}
\author[1,5]{{A.~Rocher}\orcidlink{0000-0003-4349-6424},}
\author[6]{{O.~Alves},}
\author[5]{{A.~de~Mattia},}
\author[1]{{D.~Forero-Sánchez}\orcidlink{0000-0001-5957-332X},}
\author[1,7]{{J.~Kneib},}
\author[8,9,10]{{A.~Krolewski},}
\author[11]{{T.-W.~Lan}\orcidlink{0000-0001-8857-7020},}
\author[12]{{M.~Rashkovetskyi}\orcidlink{0000-0001-7144-2349},}
\author[13]{{J.~Aguilar},}
\author[14]{{S.~Ahlen}\orcidlink{0000-0001-6098-7247},}
\author[13]{{S.~Bailey}\orcidlink{0000-0003-4162-6619},}
\author[15]{{D.~Brooks},}
\author[13]{{E.~Chaussidon}\orcidlink{0000-0001-8996-4874},}
\author[13]{{T.~Claybaugh},}
\author[16]{{A.~de la Macorra}\orcidlink{0000-0002-1769-1640},}
\author[17]{{Arjun~Dey}\orcidlink{0000-0002-4928-4003},}
\author[18]{{Biprateep~Dey}\orcidlink{0000-0002-5665-7912},}
\author[15]{{P.~Doel},}
\author[19,20]{{K.~Fanning}\orcidlink{0000-0003-2371-3356},}
\author[21,22]{{J.~E.~Forero-Romero}\orcidlink{0000-0002-2890-3725},}
\author[23,24,25]{{E.~Gaztañaga},}
\author[13]{{S.~Gontcho A Gontcho}\orcidlink{0000-0003-3142-233X},}
\author[2,26,4]{{K.~Honscheid},}
\author[27]{{C.~Howlett}\orcidlink{0000-0002-1081-9410},}
\author[17]{{S.~Juneau},}
\author[13]{{T.~Kisner}\orcidlink{0000-0003-3510-7134},}
\author[13]{{A.~Kremin}\orcidlink{0000-0001-6356-7424},}
\author[13]{{A.~Lambert},}
\author[13]{{M.~Landriau}\orcidlink{0000-0003-1838-8528},}
\author[28]{{L.~Le~Guillou}\orcidlink{0000-0001-7178-8868},}
\author[13]{{M.~E.~Levi}\orcidlink{0000-0003-1887-1018},}
\author[29,30]{{M.~Manera}\orcidlink{0000-0003-4962-8934},}
\author[2,3,4]{{P.~Martini}\orcidlink{0000-0002-4279-4182},}
\author[17]{{A.~Meisner}\orcidlink{0000-0002-1125-7384},}
\author[31,30]{{R.~Miquel},}
\author[32]{{J.~Moustakas}\orcidlink{0000-0002-2733-4559},}
\author[33]{{E.~Mueller},}
\author[16]{{A.~Muñoz-Gutiérrez},}
\author[34]{{A.~D.~Myers},}
\author[35]{{J.~Nie}\orcidlink{0000-0001-6590-8122},}
\author[36,37]{{G.~Niz}\orcidlink{0000-0002-1544-8946},}
\author[5,13]{{N.~Palanque-Delabrouille}\orcidlink{0000-0003-3188-784X},}
\author[8,9,10]{{W.~J.~Percival}\orcidlink{0000-0002-0644-5727},}
\author[13,38,39]{{C.~Poppett},}
\author[40]{{F.~Prada}\orcidlink{0000-0001-7145-8674},}
\author[41]{{M.~Rezaie}\orcidlink{0000-0001-5589-7116},}
\author[42]{{G.~Rossi},}
\author[43]{{E.~Sanchez}\orcidlink{0000-0002-9646-8198},}
\author[44]{{E.~F.~Schlafly}\orcidlink{0000-0002-3569-7421},}
\author[13]{{D.~Schlegel},}
\author[45,6]{{M.~Schubnell},}
\author[46]{{H.~Seo}\orcidlink{0000-0002-6588-3508},}
\author[17]{{D.~Sprayberry},}
\author[6]{{G.~Tarl\'{e}}\orcidlink{0000-0003-1704-0781},}
\author[17]{{B.~A.~Weaver},}
\author[28]{{P.~Zarrouk}\orcidlink{0000-0002-7305-9578},}
\author[47]{{C.~Zhao}\orcidlink{0000-0002-1991-7295},}
\author[13]{{R.~Zhou}\orcidlink{0000-0001-5381-4372},}
\author[35]{{H.~Zou}\orcidlink{0000-0002-6684-3997}}

\emailAdd{jiaxi.yu@epfl.ch}

\abstract{
Dark Energy Spectroscopic Instrument (DESI) uses more than 2.4 million Emission Line Galaxies (ELGs) for 3D large-scale structure (LSS) analyses in its Data Release 1 (DR1). Such large statistics enable thorough research on systematic uncertainties. In this study, we focus on spectroscopic systematics of ELGs. The redshift success rate ($\SSR$) is the relative fraction of secure redshifts among all measurements. It depends on observing conditions, thus introduces non-cosmological variations to the LSS. We, therefore, develop the redshift failure weight ($\wzfail$) and a per-fibre correction ($\czfail$) to mitigate these dependences. They have minor influences on the galaxy clustering. For ELGs with a secure redshift, there are two subtypes of systematics: 1) catastrophics (large) that only occur in a few samples; 2) redshift uncertainty (small) that exists for all samples. The catastrophics represent 0.26\% of the total \Yone ELGs, composed of the confusion between \oii and sky residuals, double objects, total catastrophics and others. We simulate the realistic 0.26\% catastrophics of \Yone ELGs, the hypothetical 1\% catastrophics, and the truncation of the contaminated $1.31<z<1.33$ in the \abacus ELG mocks. Their $P_\ell$ show non-negligible bias from the uncontaminated mocks. But their influences on the redshift space distortions (RSD) parameters are smaller than $0.2\sigma$. The redshift uncertainty of \Yone ELGs is 8.5$\kms$ with a Lorentzian profile. The code for implementing the catastrophics and redshift uncertainty on mocks can be found in \url{https://github.com/Jiaxi-Yu/modelling_spectro_sys}. 
}

\begin{document}
\maketitle
\flushbottom

\section{Introduction} 
\label{sec:intro}
Galaxy redshift surveys probe the 3D LSS of the Universe by measuring the redshifts of millions of galaxies and quasars. Baryonic acoustic oscillations (BAO;~\cite{1998bao}) and redshift-space distortions (RSD;~\cite{1987rsd}) encoded in the galaxy clustering reflect the properties of dark energy and dark matter. However, artefacts, or observational systematics, may introduce biases or reduce the precision of the cosmological measurements via BAO and RSD \citep[e.g.,][]{sys2010blunder,sys2017boss,eBOSS_ELG_2021,eBOSS_ELG_RSD,eBOSS_ELG_Fourier_2021}. Therefore, identifying, describing, and correcting any systematics that may exist in the cosmological measurements is essential for redshift surveys.

Dark Energy Spectroscopic Instrument (DESI, 2021-2026) \cite{Levi2013,DESICollaboration2016a,DESICollaboration2016b} is conducting the largest galaxy redshift survey to date with the 4-meter Mayall Telescope at Kitt Peak, Arizona, US. DESI has already observed thousands of square degrees of the sky using robotic-controlled fibres \cite{DESI2022.KP1.Instr}. It is on track to obtain the spectra of $\sim 40$ million galaxies and quasars over $14,000\degsq$ of the sky. Its early data release in June 2023 (EDR\footnote{\url{https://data.desi.lbl.gov/doc/releases/edr}}, \cite{DESI2023b.KP1.EDR}) published the data collected during the survey validation phase \cite{DESI2023a.KP1.SV} from December 2020 to May 2021, covering several hundred square degrees of sky. Its 5-year survey then started in May 2021, and the data collected during the first year of observations will be made available in the first data release (\Yone, \cite{DESI2024.I.DR1}). \Yone uses 300,017 Bright Galaxies between redshift $0.1<z<0.4$ (BGS; \cite{BGS.TS.Hahn.2023}), 2,138,600 Luminous Red Galaxies between redshift $0.4<z<1.1$ (LRGs; \cite{LRG.TS.Zhou.2023}), 2,432,022 Emission Line Galaxies between redshift $0.8<z<1.6$ (ELG; \cite{ELG.TS.Raichoor.2023}) and 1,223,170 quasars (QSO) between redshift $0.8<z<3.5$ \cite{QSO.TS.Chaussidon.2023} that are observed over more than $7400\degsq$ of the sky for cosmological analysis \cite{DESI2024.II.KP3}. These dark matter tracers are grouped into LSS catalogues with different weights \cite{KP3s15-Ross}, aiming at providing unbiased clustering of galaxies, quasars, and the Ly$\alpha$ forest \cite{DESI2024.II.KP3, DESI2024.IV.KP6}. Accurate estimations of the full covariance matrices \cite{KP3s8-Zhao,KP3s12-Prada,KP4s8-Alves,KP4s7-Rashkovetskyi,KP4s6-Forero-Sanchez} and data blinding \cite{KP3s9-Andrade} are also crucial for the robustness of the cosmological results. The percent-level precision measurements of BAO are presented in \cite{DESI2024.III.KP4,DESI2024.IV.KP6}, where we show that the BAO precisions of the 1-year DESI observation outperform those from the two-decadal Sloan Sky Digital Survey (SDSS\footnote{\url{https://www.sdss4.org/science/final-bao-and-rsd-measurements/}}; \cite{SDSS_cosmology2021}). The RSD measurements of DR1 will come very soon in \cite{DESI2024.V.KP5}. The cosmological constraints on $\Omega_m$, $H_0$, $\Omega_K$, $M_\nu$ and dark energy equation of state informed by BAO are concluded in \cite{DESI2024.VI.KP7A}, presenting a $3\sigma$ detection of a $w_0w_a$CDM cosmology compared with the $\Lambda$CDM. The cosmological constraints from RSD and the measurements of primordial non-Gaussianity $f_{\rm NL}$ will be introduced in \cite{DESI2024.VII.KP7B,DESI2024.VIII.KP7C}. 

For DESI DR1, we consider three types of observational systematics that could influence the cosmological measurements: target selection \cite{TS.Pipeline.Myers.2023}, fibre assignment \cite{FBA.Raichoor.2024} and redshift measurements \cite{Spectro.Pipeline.Guy.2023,Redrock.Bailey.2024}. DESI Legacy Imaging Survey \cite{LS.dr9.Schegel.2024} was built to provide galaxy and quasar candidates for spectroscopic observations. The Beijing-Arizona Sky Survey (G, R bands), and the Mayall Z-band Legacy Survey (\DN; \cite{BASS.Zou.2017}) comprise the north galactic cap of the DESI footprint with declination larger than $32.375^\circ$. The rest of the DESI footprint is covered by the Dark Energy Camera Legacy Survey (DECaLS; \cite{LS.Overview.Dey.2019}) and the Dark Energy Survey (DES; \cite{DES2005}) observed with the Dark Energy Camera (DECam; \cite{DECam2015}). The systematics analyses of the redshift measurements are thus split into \DN and \DS parts due to the difference in their instruments. Imaging systematics are artefacts introduced from the target selection by, for example, the accuracy of the photometric information and the selection criteria. The modelling and corrections of imaging systematics are described in detail in \cite{KP3s2-Rosado,KP3s13-Kong,KP3s14-Zhou,KP3s10-Chaussidon}. Fibre assignment with proper designs ensures that the redshift survey will be completed on time but will also introduce unphysical variations in the observed galaxy clustering. \cite{KP3s6-Bianchi,KP3s7-Lasker,KP3s11-Sikandar,KP3s5-Pinon} address the effects of it from different aspects. \cite{KP3s3-Krolewski} and this paper characterize the impact of bad redshift measurements for all types of tracers. Specifically, this paper focuses on ELGs, the largest group of objects in the DESI survey, due to their unique role in probing the LSS in the star-forming epoch at $1<z<2$. 

%\jiaxi{$\sim30\%$ of the ELG targets did not receive good redshift measurements \cite{ELG.TS.Raichoor.2023}, much larger than the other tracers (e.g., 1.2\% for \Yone LRGs \cite{KP3s3-Krolewski}). Therefore, the weight to make up for the loss of the ELG targets should be different from the others \cite{2018_eBOSSwzfail}.} Moreover, t
The success rate of the redshift measurement ($\SSR$) varies with observing conditions, reflecting the unphysical galaxy density variations brought by the instruments of DESI. Correcting these effects requires fair up- or down-weighting on galaxy samples with secure redshift measurements, thus avoiding potential impact on the LSS analysis. For those with \textit{secure} redshift measurements, their redshift are not necessarily the \textit{true} redshift. A small fraction of ELGs, e.g., $\sim$0.3\% for eBOSS DR16 \cite{eBOSS_ELG_2021} and DESI preliminary studies \cite{ELG.TS.Raichoor.2023,VIGalaxies.Lan.2023}, have \textit{secure} redshifts that are very different from its \textit{true} redshift for various reasons. This may also introduce bias to the clustering measurements by significantly altering the LSS. In most cases, \textit{secure} redshift measurements are close to its \textit{true} redshift, with statistical uncertainties. It is as small as $10\kms$ for DESI ELGs \cite{ELG.TS.Raichoor.2023,mine}, resulting in negligible clustering impact on $s>5\mpch$. We will study all three aspects of the spectroscopic systematics for \Yone ELGs at $0.8<z<1.6$ and quantify their clustering and cosmological impacts.

This paper is arranged as follows. We describe the \Yone data, especially the corrections on the redshift success rate, galaxy mocks, and covariance matrices involved in our systematics analysis in \cref{sec:data}. The two-point clustering estimators and DESI cosmological pipeline will be introduced in \cref{sec:method}. \cref{sec:weights} presents the characterization and corrections of the redshift failures. The feature and impact of redshift catastrophics and uncertainty of ELGs with reliable redshift measurements will be introduced in \cref{sec:catas}. Finally, we conclude our findings in \cref{sec:conclusion}. 

In this work, we adopt a flat $\Lambda$CDM fiducial cosmology from the mean results of Planck \cite{Planck2018} \textsc{base\_plikHM\_TTTEEE\_lowl\_lowE\_lensing}  with $\omega_{\mathrm{b}} = 0.02237, \; \omega_{\mathrm{cdm}} = 0.12, \; h = 0.6736, \;  A_{\mathrm{s}} = 2.083 \cdot 10^{-9}, \; n_{\mathrm{s}} = 0.9649, \; \sigma_8=0.8079, \; N_{\mathrm{eff}} = 3.044, \; \sum m_{\nu} = 0.06 \; \mathrm{eV}$ (single massive neutrino). This is the fiducial DESI cosmology when converting the positions of objects to Cartesian coordinates in the two-point statistics computation, and also the cosmology of ELG mocks \cite{antoine2023} built on \abacus simulations \cite{AbacusSummit}.

\section{Data and Mocks} 
\label{sec:data}
\subsection{DESI \Yone Data} 
\label{sec:desiy1}
The \Yone data of DESI is the first year of the DESI Main Survey (May 2021 to June 2022, \cite{DESI2024.I.DR1}). Covering over 7,400 square degrees of the sky, \Yone includes more than 6 million spectra of galaxies and quasars with accurate redshift at $0.1<z<3.5$ for cosmological measurements with LSS. We focus on the spectroscopic systematics of \texttt{ELG\_LOPnotqso} samples (\elg hereafter) in DESI \Yone \cite{KP3s15-Ross}. This corresponds to ELG targets that have a high observational priority (\texttt{ELG\_LOP} with a surface number density $\approx 1940\,$deg$^{-2}$) and are not part of QSO targets ($\approx5\%$ of \texttt{ELG\_LOP}) \cite{ELG.TS.Raichoor.2023}. We include three types of data products in our study: 1) the \texttt{full\_HPmapcut} catalogue that includes all \Yone \elg targets with a good observing prerequisite and imaging properties \cite{DESI2024.II.KP3}. It means that all observed ELG targets (not necessarily true ELGs) with good and failed redshift measurements at all redshifts are included. This catalogue is used to construct $\wzfail$ (\cref{sec:wzfail}) and explore the possible improvement of it (\cref{sec:ssrcorr}). We refer to it as \full catalogue afterwards; 2) the \Yone \elg LSS catalogues with ELGs at $0.8<z<1.6$ \cite{KP3s15-Ross}. It includes the total observational systematics weight $w_{\rm tot}$, assuming systematics are decomposable and all corrections are non-overlapping. $w_{\rm tot}$ is obtained as follows \citep{KP3s15-Ross,DESI2024.II.KP3}:
\begin{equation}
    w_{\rm tot} = w_{\rm comp}w_{\rm sys}\wzfail ,
    \label{eq:wtot}
\end{equation}
where $w_{\rm comp}$ is for the correction of the target completeness due to fibre assignment, $w_{\rm sys}$ is to correct the target density fluctuation existing in the \DN and \DS imaging surveys (see \cite{DESI2024.II.KP3} for more details), and $\wzfail$ is for the failed-redshift systematics (see \cref{sec:wzfail}). The LSS catalogue of ELGs is used to quantify the clustering impact of the redshift failure weight $\wzfail$ and its corrections; 3) spectroscopically confirmed ELGs from the One-Percent Survey of EDR. This dataset will be used to construct a catalogue of repeated observation (\cref{sec:repeats}) for systematics studies in \cref{sec:catas} and provide the visual inspection of catastrophics in \cref{appendix:catas-spec}.

\subsubsection{Redshift Success Rate \texorpdfstring{$\SSR$}{ssr}} 
\label{sec:ssr}
$\wzfail$ are weights that account for missing objects in the observed LSS (i.e., not in LSS catalogues) due to failed redshift measurements. An ELG redshift measurement is classified as successful if it meets the requirement as follows \cite{ELG.TS.Raichoor.2023}
\begin{equation}
    {\rm log_{10}}\left(S_{\rm \oii}\right)+0.2{\rm log_{10}}(\Delta\chi^2)>0.9,
    \label{eq:good elg}
\end{equation}
where $(S_{\rm \oii})$ is the signal-to-noise ratio (SNR) of the \oii emission line fit, and $\Delta\chi^2$ is the $\chi^2$ difference between the best-fit redshift and the second-best redshift in the \textsc{Redrock} pipeline. This means that a reliable ELG redshift measurement should have either a high \oii SNR or a large $\Delta\chi^2$. %a clear \oii feature (i.e.,

The success rate of redshift measurement is the complement of the redshift failure rate. We always use the normalized redshift success rate $\SSR$ in this study, defined as
\begin{equation}
    \SSR (x) = \frac{N_{\rm goodz}(x)/N_{\rm obs}(x)}{\sum^xN_{\rm goodz}(x)/\sum^xN_{\rm obs}(x)}, 
    \label{eq:define ssr}
\end{equation}
where $x$ represents different observing conditions (e.g., the effective observing time, the position on the focal plane). $N_{\rm obs}$ is the number of \elg targets observed in appropriate conditions with no instrumental issue, and $N_{\rm goodz}$ is a subsample of $N_{\rm obs}$ with good redshift measurements (\cref{eq:good elg}). For \Yone ELGs, $\sum^xN_{\rm goodz}(x)/\sum^xN_{\rm obs}(x)= 72.6\%$ \cite{KP3s3-Krolewski}. 

We remind our readers that we measure $\SSR$ for \DN and \DS footprint separately. It means that all $\SSR$ studies and discussions are divided into these two areas (i.e., area selections are applied on both $N_{\rm goodz}$ and $N_{\rm obs}$). Additional selections (e.g., on redshift range as discussed in \cref{fig:ssr_correction_dz0.1}) should be implemented on $N_{\rm goodz}(x)$ only as they have reliable properties including redshift measurements. 

\subsubsection{Repeated Observation Catalogue} 
\label{sec:repeats}
The studies of redshift catastrophics and uncertainty are based on repeated observations of the same object. The DESI One-Percent Survey, part of the survey validation program \cite{DESI2023a.KP1.SV}, has a footprint that overlaps with the \Yone footprint. In DR1, the same object can be observed repeatedly on different nights. So, we select all repeated redshift measurements of \texttt{ELG\_LOPnotqso} samples (\Yone ELGs hereafter) by cross-match these two data sets processed with the same versions of \textsc{Redrock} \cite{Redrock.Bailey.2024}, and we compute the redshift difference $\Delta v = \Delta z c/(1+z)$ iteratively among pairs. Studies based on these repeated observations represent the properties of DESI \Yone ELGs since they are fair subsamples by construction (see, e.g. \cref{sec:catas_feature})

There are 115,160 pairs of repeated observation for \Yone ELGs, and 307 pairs (0.26\%) of them have $|\Delta v|>1000\kms$, which are the catastrophics in our study. This is similar to the ELG catastrophics rate of the DESI survey validation data \cite{ELG.TS.Raichoor.2023}. %and that of eBOSS \cite{eBOSS_ELG_2021}. 
We will provide a detailed description of these samples in \cref{sec:catas_feature}. Pairs with $|\Delta v|<1000\kms$ will be used to study the redshift uncertainty (\cref{sec:zerr}). $\Delta v$ is roughly symmetric w.r.t. 0 as illustrated in \cref{sec:zerr}. 

\subsection{Galaxy Mocks and Covariance Matrices} 
\label{sec:abacusmocks}
We employ galaxy mocks that mimic the \Yone ELG samples to assist our study of spectroscopic systematics. The model ELGs were generated by implementing a modified high-mass-quenched HOD model \cite{antoine2023} on \abacus simulations \cite{AbacusSummit} in a $2\gpch$ box at $z=1.100$. Next, these model galaxies in simulation boxes were downsampled to the observed redshift distribution $n(z)$ of \Yone \elg samples and truncated to a spherical shell that matches the footprint of the DESI \Yone survey. The survey-like model ELGs then went through the data reduction pipeline of real observations \cite{KP3s7-Lasker} (i.e., \texttt{altmtl}) to select the `observed' ELGs. 
%\jiaxi{a fast fibre-assignment (FFA) process \cite{KP3s11-Sikandar}}  
%\jiaxi{I mixed catastrophics in this step already, so the number of ELGs in this catalog is not so meaningful} 
The output mock ELG catalogue is similar to the \full catalogue as mentioned in \cref{sec:desiy1}, and we will apply the modelled spectroscopic systematics (see \cref{sec:catas}) to this catalogue. Finally, mocks with and without systematics will be processed by the pipeline to generate LSS catalogues for data (data type 2 in \cref{sec:desiy1}) and to calculate the two-point statistics (see \cref{sec:2pt}) for comparison. There are 25 realizations of \abacus simulations. Therefore, we will implement the systematics on all 25 realizations of the ELG mocks, and their averaged clustering will be used for cosmological tests to reduce the effect of cosmic variance.

The cosmological measurements require accurate estimations of the overall cosmic variance, i.e. the full covariance matrices. We also need accurate covariance matrices to quantify the clustering and cosmological impact of spectroscopic systematics. In our study, we use analytical covariances created by 
\textsc{thecov}\footnote{\url{https://github.com/cosmodesi/thecov}} \cite{KP4s8-Alves} in the Fourier space for the power spectrum multipoles, $P_\ell(k)$, which is based on \textsc{CovaPT}\footnote{\url{https://github.com/JayWadekar/CovaPT/}} \cite{CovaPT_Wadekar:2019rdu} 
%\jiaxi{\textsc{EZmock}\footnote{\url{https://github.com/cheng-zhao/pyEZmock}} \cite{EZmock2015,EZmock2021,KP3s8-Zhao} fast galaxy mocks in Fourier space} 
and \textsc{RascalC}\footnote{\url{https://github.com/oliverphilcox/RascalC}}\footnote{\url{https://github.com/misharash/RascalC-scripts/}} in the configuration space for the two-point correlation function (2PCF) multipoles, $\xi_\ell(s)$ \cite{2023MNRAS.524.3894R,KP4s7-Rashkovetskyi}. We direct the reader to \cite{KP4s6-Forero-Sanchez} for a detailed study of the covariance matrices in DESI DR1.

\section{Method} 
\label{sec:method}
\subsection{Two-point Statistics} 
\label{sec:2pt}
Two-point statistics describe the probability of finding excess pairs of galaxies compared to a random distribution. The detailed description of the computation in DESI \Yone can be found in \cite{DESI2024.II.KP3}. In this section, we provide a summary of their computational techniques.

The distribution of galaxies from LSS catalogues has been corrected by $w_{\rm tot}$ to screen the impact of observational systematics. In addition, the evolution of the galaxy number density as redshift will introduce extra clustering variance at BAO scale \citep{FKP1994}. We thus include the FKP weights $w_{\rm FKP}$ in the clustering measurement, defined as 
\begin{equation}
    w_{\rm FKP} = \frac{1}{1+n_\mathrm{local}(z)P_0},
\end{equation}
where $n_\mathrm{local}(z)$ is the tracer average number density at redshift $z$, and $P_0= 4000\, h^{-3}\,{\rm Mpc}^{3}$ for ELGs, is the amplitude of the observed power spectrum at $k_0 \approx 0.15\invmpch$. Therefore, we weight each galaxy with $w_{\rm tot}w_{\rm FKP}$ for 2-point statistics measurements.

In the configuration space, we measure the two-point correlation function $\xi$ (2PCF). It depends on the comoving distance $s$ between pairs of galaxies and $\mu$, which is the cosine of the angle between the distance vector $\boldsymbol{s}$ and the line-of-sight. For galaxies observed from redshift surveys, we use the Landy--Szalay estimator (LS) to calculate the $\xi$ \cite{Landy1993}: 
\begin{equation}
\label{LS estimator}
   \xi_{\rm LS}  = \frac{\rm DD-2DR+RR}{\rm RR}, 
\end{equation}
where DD, DR and RR represent the number of galaxy pairs identified in data--data catalogues, data--random catalogues, and random--random catalogues at a given distance and angle $(s,\mu)$ normalized by their corresponding total number of pairs in these catalogues. $\xi(s,\mu)$ can be decomposed by Legendre polynomials $L_\ell(\mu)$ to obtain the multipoles of $\xi$ as 
\begin{equation}
    \label{multipoles}
    \xi_\ell (s) = \frac{2\ell+1}{2}\int_{-1}^1 \xi(s,\mu) L_\ell(\mu) {\rm d}\mu.
\end{equation}
We compute the galaxy pairs for observations using the DESI package \textsc{pycorr}\footnote{\url{https://github.com/cosmodesi/pycorr}}, a wrapper of the \texttt{CORRFUNC} package \cite{pycorr2020,pycorr_corrfunc}, between 0-200$\mpch$ in 200 linear bins and with 200 linear $\mu$ bins and regroup them to obtain the $\xi_\ell$ with $\Delta s=4\mpch$ for $\ell=0,2,4$. We use the comoving distance at $s\in[30,200]\mpch$ for $\chi^2$ calculations (\cref{sec:chi2-comparison}).

Power spectra $P(k)$ are two-point statistics in Fourier space. Its estimator \cite{yamamoto2006} is based on the weighted field \cite{FKP1994}:
\begin{equation}
F(\boldsymbol{r}) = n_{d}(\boldsymbol{r}) - \alpha n_{r}(\boldsymbol{r}).
\label{eq:fkp}
\end{equation} 
where $\boldsymbol{r}$ is the 3D Cartesian coordinate, $n_d$ and $n_r$ are the binned $w_{\rm tot}$-weighted data and $w_{\rm tot}$-weighted random catalogues on a grid of cell size 6$\mpch$, and $\alpha=\sum^{N_d}_{i=1}w_{\rm tot,i(d)}/\sum^{N_r}_{i=1}w_{\rm tot,i(r)}$ rescales the mean density of the random catalogue to the mean data density. The power spectrum multipoles $P_\ell(k)$ can be written as 
\begin{equation}
P_{\ell}(k) = \frac{2 \ell + 1}{A N_{k}} \sum_{\boldsymbol{k}} F_{0}(\boldsymbol{k}) F_{\ell}(-\boldsymbol{k}) - \mathcal{SN}_{\ell}
\label{eq:power_spectrum_multipoles}
\end{equation}
where $\boldsymbol{k}$ is the wavenumber between galaxy pairs and $N_k$ the number of modes in a $\boldsymbol{k}$ bin, and 
\begin{equation}
F_{\ell}(\boldsymbol{k}) = \sum_{\boldsymbol{r}} F(\boldsymbol{r}) \mathcal{L}_{\ell}(\hat{\boldsymbol{k}} \cdot \hat{\boldsymbol{r}}) e^{i \boldsymbol{k} \cdot \boldsymbol{r}}.
\label{eq:fkp_multipoles}
\end{equation}
where $\hat{\boldsymbol{k}} \cdot \hat{\boldsymbol{r}}$ is the cosine of the angle between the wavenumber vector $\boldsymbol{k}$ and the line-of-sight $\hat{\boldsymbol{r}}$. The shot-noise $\mathcal{SN}_{\ell}$ is non-zero for $\ell=0$ as
\begin{equation}
\mathcal{SN}_{0} = \frac{1}{A} \left[\sum^{N_d}_{i=1}w_{\rm tot,i(d)}^{2}+\alpha^2\sum^{N_r}_{i=1}w_{\rm tot,i(r)}^{2} \right].
\end{equation}
The normalization term $A=\alpha / dV \sum^k n_{d,k}n_{r,k}$ is summed over a cell with $dV^{1/3} = 10 \; \mpch$ in size. We compute our observed $P_\ell(k)$ with \textsc{pypower}\footnote{\url{https://github.com/cosmodesi/pypower}} \cite{pypower2017} from 0--0.4$\invmpch$ with $\Delta k=0.005$ for $\ell=0,2,4$. The $\chi^2$ calculations include $k\in[0.02,0.35]\invmpch$ (\cref{sec:chi2-comparison}).

\subsection{Quantifying the Clustering Differences} 
\label{sec:chi2-comparison}
We use $\chi^2_{\rm sys}$ to describe the differences between the standard clustering and the clustering for mocks with spectroscopic systematics, defined as
\begin{equation}
\chi_{\rm sys}^2  =  (\boldsymbol{F}_{\ell, \rm std}-\boldsymbol{F}_{\ell, \rm sys})^T\mathbf{C}^{-1}(\boldsymbol{F}_{\ell, \rm std}-\boldsymbol{F}_{\ell, \rm sys}),
\label{eq:chi2_general}
\end{equation}
where $\boldsymbol{F}$ denotes the vector composed of the two-point statistics multipoles. We calculate the $\chi_{\rm sys}^2$ values for $\boldsymbol{F}_{\ell=0,2,4}=(\xi_0,\xi_2,\xi_4)$ at $30<s<200\mpch$ in the configuration space and $\boldsymbol{F}_{\ell=0,2,4}=(P_0,P_2,P_4)$ at $0.02<z<0.35\invmpch$ in Fourier space.
%We calculate the $\chi_{\rm sys}^2$ values for the reference of BAO measurement impact with $\boldsymbol{F}_{\ell=2}=(\xi_0,\xi_2)$ in the configuration space and $\boldsymbol{F}_{\ell=2}=(P_0,P_2)$ in Fourier space. $\boldsymbol{F}_{\ell=4}=(\xi_0,\xi_2,\xi_4)$ and $\boldsymbol{F}_{\ell=4}=(P_0,P_2,P_4)$ are used to calculate $\chi_{\rm sys}^2$ for the Full-Shape RSD measurement. 
$\boldsymbol{F}_{\ell, \rm std}$ represents the standard clustering measurement which is 1) ELGs from LSS catalogues with $w_{\rm tot}$ (\cref{tab:SSR_chi2} from \cref{sec:weights}) and 2) \abacus ELG mocks without catastrophics (\cref{tab:catas_chi2} from \cref{sec:catas}). $\boldsymbol{F}_{\ell, \rm sys}$, correspondingly, is the clustering of 1) ELGs with other redshift success corrections and 2) \abacus ELG mocks with different catastrophics. $\mathbf{C}^{-1}$ is the inverse of the analytical covariance matrices. $\sqrt{\chi_{\rm sys}^2}$ estimates the maximum deviations the systematics can introduce to cosmological measurement rescaled by the cosmic variance. We think the effect of a type of systematics is negligible if $\sqrt{\chi_{\rm sys}^2}<1$ (i.e., cosmological impacts $<1\sigma$). To avoid confusion, we use $\varepsilon_\xi$ and $\varepsilon_P$ for the 1$\sigma$ error of the galaxy clustering in the configuration space and Fourier space, respectively in the following sections. They are the square root of the diagonal terms of the analytical covariance matrices $\varepsilon_i=\sqrt{C_{i,i}}$.

\subsection{RSD Tests: ShapeFit and Full Modelling}
\label{sec:desilike}
We study the influence of ELG spectroscopic systematics on RSD measurements with full modelling method and ShapeFit compression \cite{RSDtheory_velocileptors,RSDtheory_FOLPSnu,RSDtheory_pybird,RSDtheory_configspace}. The full modelling of RSD is to generate model-dependent theoretical power spectra $P_{\rm theory}$ with linear power spectra $P_{\rm lin}$ provided directly by Boltzmann codes such as \textsc{CLASS} \cite{CLASS} and \textsc{CAMB} \cite{CAMB}. This algorithm is thus computationally expensive, since each sampling of cosmology requires the calculation of the accurate $P_{\rm lin}$. ShapeFit compression, in contrast, is a more efficient, model-independent way of producing $P_{\rm theory}$ from $P'_{\rm lin}$. The $P'_{\rm lin}$ of ShapeFit is parameterized as follows \cite{ShapeFit2021}
\begin{equation}
P'_{\rm lin}(k) = P_{\rm lin}^{\rm fid}(k) {\rm exp}\left\{ \frac{m}{a}{\rm tanh} \left[ a\text{ln}\left( \frac{k}{k_p} \right) \right] \right\}, 
\end{equation}
where $P_{\rm lin}^{\rm fid}$ is the linear power spectra under the DESI fiducial cosmology, $k_p=\pi /r_d$ is the pivot scale and $r_d$ is the sound horizon scale, $a=0.6$. $m$ is the free parameter to approximate the accurate linear power spectra. In our tests, we use \textsc{velocileptors}\footnote{\url{https://github.com/sfschen/velocileptors}} for both tests wrapped in \textsc{desilike}\footnote{\url{https://github.com/cosmodesi/desilike}}, the DESI cosmological pipeline. To speedup the fitting, we employ the emulator in \textsc{desilike} based on the Taylor expansion . The cosmological parameters for full modelling are $\Theta=\{h,\;\omega_{\rm cdm},\;\omega_b,\;\As \}$ where $\omega_*=\Omega_*h^2$. For ShapeFit $\Theta=\{q_{\rm iso},\;q_{\rm AP},\;df,\;dm\}$, representing the isotropic and anisotropic BAO dilation (see \cite{DESI2024.III.KP4} for detailed explanations), $df = f/f^{\rm fid}$ is the difference between the measured linear growth rate $f$ and the fiducial value, $dm=m-1$.
 
To take into account the observational geometry, we need to apply the window matrix $W_{\tilde{\ell},\ell}$ on $P_{\ell,\rm theory}$ as $\tilde{P}_{\tilde{\ell},\rm theory}(\tilde{k}_i)=\sum_j W_{\tilde{\ell},\ell}(\tilde{k}_i,k_j)P_{\ell,\rm theory}(k_j)$. Assuming a Gaussian likelihood $L(\Theta)={\rm exp}(-\chi^2(\Theta)/2)$, where $\chi^2$ is defined as 
\begin{equation}
\chi^2  =  (\boldsymbol{P}_{\ell, \rm mock}-\tilde{\boldsymbol{P}}_{\tilde{\ell}, \rm theory})^T\mathbf{C}^{-1}(\boldsymbol{P}_{\ell, \rm mock}-\tilde{\boldsymbol{P}}_{\tilde{\ell}, \rm theory}),
\label{eq:chi2_cosmo}
\end{equation}
$\boldsymbol{P}_{\ell=0,2} = (P_0,P_2)$ represent the data vector of systematics-uncontaminated and contaminated mocks, $\tilde{\boldsymbol{P}}_{\ell=0,2} = (\tilde{P}_0,\tilde{P}_2)$ is the data vector of geometry-modulated theory. We consider $0.02<k<0.2\invmpch$ in the RSD fitting.

\section{Redshift Success Rate Corrections}
\label{sec:weights}
\subsection{The Redshift Failure Weight \texorpdfstring{$\wzfail$}{wzfail}} 
\label{sec:wzfail}
% the definition of wzfail of ELG, special, 
%% https://github.com/Jiaxi-Yu/LSS/blob/main/py/LSS/ssr_tools_new.py#L149
%% https://github.com/Jiaxi-Yu/LSS/blob/main/py/LSS/main/cattools.py#L3264
Redshift failures are observed objects that do not have reliable redshift measurements. They are dropped from the LSS catalogue and thus may lead to an underestimation of galaxy density in the observed 3D map of the Universe. A straightforward way of making up for the absence of objects is to up-weigh the nearest object to the failed observation on the sky, i.e., count it as two objects when calculating the galaxy clustering. This simple weighting scheme was used in LRGs from the Baryon Oscillation Spectroscopic Survey (BOSS, 2008--2014; \cite{BOSS}) from SDSS-III \cite{SDSSIII}. It worked well as the failed redshift measurements were distributed randomly on the focal plane and only composed of less than 2\% of the total observation \cite{boss_zfail_2012}. 

%\jiaxi{Will's comment (resolved): An upweight of the good observations to correct for the bad ones will only work where the selection is probabilistic. I.e. the unobserved and observed galaxies are drawn randomly from the same sample with some probability. This will work where the failure fraction is small (and random), but must fail for large failure fractions - "some galaxy types will never get a good redshift in some regions."=> not true;}
%\jiaxi{Will's comment(resolved): If the upweighted population is intrinsically different (e.g. different bias) from the populations with good redshifts, there is a problem. Worth discussing this. }

However, when the failure rate rises to 10\% or higher, as seen with LRGs from the extended Baryon Oscillation Spectroscopic Survey (eBOSS, 2014--2020; \cite{Dawson2015}) of SDSS-IV \cite{SDSSIV} and ELGs from eBOSS and DESI \cite{eBOSS_ELG_2021,ELG.TS.Raichoor.2023}, such a simple up-weighting approach becomes ineffective. This is because it assumes that the properties of galaxies with failed redshifts are similar to those of the successful samples, which is often too strong  (see \cite{ELG.TS.Raichoor.2023} for the example of ELGs). In such cases, the goal of correcting failed redshift measurements should be to restore the uniformity of the redshift success rate across varying observing conditions rather than striving for a 100\% success rate. Similar to \cite{2018_eBOSSwzfail}, we model the observed redshift success rate $\SSR$ as a function of observing conditions and use it to construct the $\wzfail$ weight. To simplify the $\SSRmodel$, we normalize the $\SSR$ as shown in \cref{eq:define ssr}.

%\jiaxi{This is because a galaxy distribution with a constant failure rate has the same density contrast, thus clustering as a distribution without a failure. Moreover, correcting a non-unity success rate back to unity assumes that the failed samples have the same properties as the success ones (e.g., the redshift distribution), which is not always the case \cite{ELG.TS.Raichoor.2023}.} So, our $\wzfail$ correction is designed to recover consistent redshift success rates regardless of the observing conditions, i.e., to remove the influence of the spectroscopic observation on the intrinsic galaxy distribution. 

The primary observing condition influencing $\SSR$ is the observational squared SNR of ELG spectra \texttt{TSNR2\_ELG} ($\tsnr$ hereafter), which is proportional to the effective observing time. $\tsnr$ is defined as follows \cite{Spectro.Pipeline.Guy.2023}
\begin{equation}
    \tsnr = \sum_i T_i^2 \langle (\delta F)^2 \rangle_i\sigma_i^{-2}\, ,
    \label{eq:tsnr}
\end{equation}
where $T$ is the calibration coefficient of photon-electron conversion measured from each observed spectrum, involving the product of throughput\footnote{Figure 27 of \cite{DESI2022.KP1.Instr} present the throughput of DESI instrument.} (the system's efficiency of collecting incoming photons) and the real exposure time. $\langle (\delta F)^2 \rangle$ is the squared spectral line flux averaged over all spectral templates. $\sigma^2$ is the noise variance, including the instrumental noise and sky spectra. $i$ is the wavelength index of the whole spectrum. All else held constant, $\SSR$ should be monotonic with $\tsnr$. 

In addition to $\tsnr$, we find that $\SSR$ also varies as the redshift. This is because the sky emission lines appear at distinct wavelengths corresponding to distinct redshifts at which \oii emission can be observed (see Section 7.2 of \cite{ELG.TS.Raichoor.2023} for example). The $\SSR$ at these redshifts are lower than the others because the noise level (the flux of sky-spectrum residuals) is high for \oii detections. %\jiaxi{You say "we find that the redshift of ELG is the second most important variable”, but then there’s an argument that the observed wavelength is important because of sky-lines. Is it the redshift, or the observed wavelength that is most important?} 

We aim to obtain a redshift failure weight $\wzfail$ that corrects the $\SSR$ dependences on $\tsnr$ and redshift. This correction is part of the total weight $w_{\rm tot}$ in \Yone ELG LSS catalogue as introduced in \cref{sec:desiy1}. But $\wzfail$ is constructed for all observed ELG targets, including those without reliable redshift measurements. Bearing this in mind, it is necessary to find quantities other than the `redshift' itself to embody the challenges of \oii detection in the presence of sky emission lines. 
%The admittedly convoluted process we adopt for DESI \Yone ELGs is as follows.
%We wish to normalize the overall redshift success rate at every redshift so that redshifts with low success rates are not unduly up-weighted in clustering measurements. 
%(Other studies, e.g., for luminosity functions, likely prefer to use the absolute redshift dependence.) 

\subsubsection{\texorpdfstring{\oii}{OII} Emission and Redshift Measurement}
\label{sec:oii-flux-for-redshift}
\begin{figure}
\centering
\includegraphics[width=\textwidth]{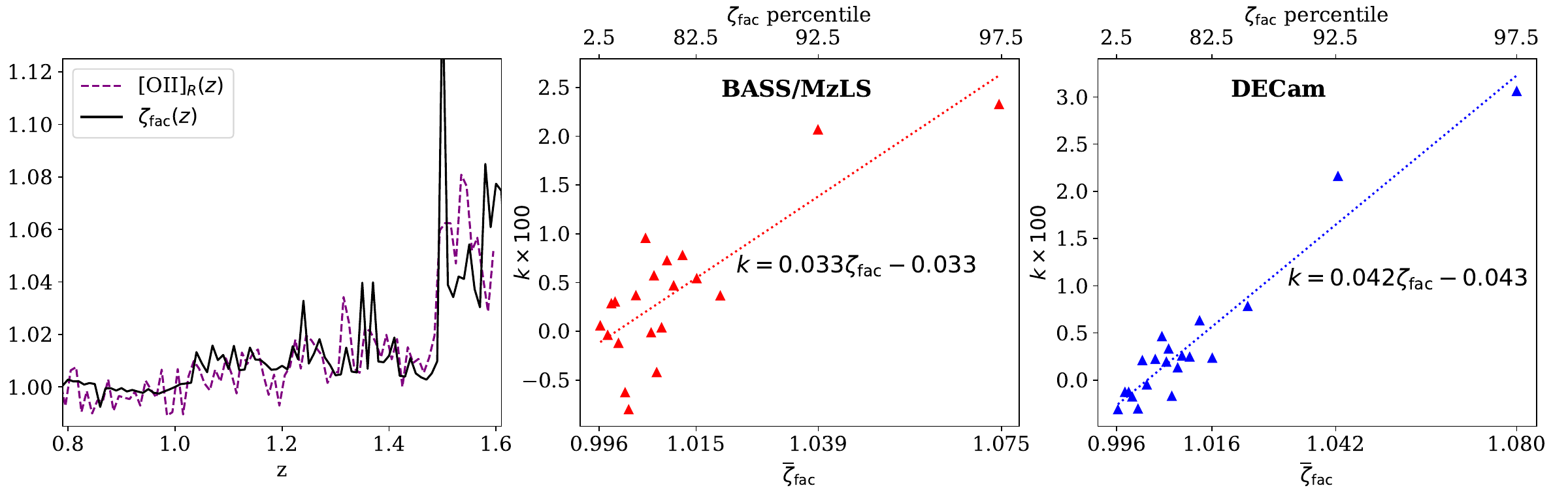}
%\caption{\textit{Left:} [O\,\textsc{ii}]$_R$ (the dashed line) and $\zeta_{\rm fac}$ (the solid line) as a function of the redshift. $\zeta_{\rm fac}(z)$ (from spectra template) is a good representation of the observed $\oiirel$, describing the difficulty (the larger they are, the more difficult) of obtaining a good redshift measurement. \textit{Right:} The 20 slopes $k$ and their corresponding $\overline{\zeta}_{\rm fac}$ obtained from \DN (triangles) and \DS (stars) and their best-fitting linear functions (the dotted line for \DN and the dashed line for \DS) for \Yone \elg. 
\caption{\textit{Left:} [O\,\textsc{ii}]$_R$ (the dashed line) and $\zeta_{\rm fac}$ (the solid line) as a function of the redshift. $\zeta_{\rm fac}(z)$ (from spectra template) is a good representation of the observed $\oiirel$, describing the difficulty (the larger they are, the more difficult) of obtaining a good redshift measurement. \textit{Middle:} The 20 slopes $k$ and their corresponding $\overline{\zeta}_{\rm fac}$ (the bottom x-label) and $\zeta_{\rm fac}$ percentiles (the top x-label) for \DN. We implement linear regression on the $k$-$\overline{\zeta}_{\rm fac}$ relation and the best-fit relation is shown in the dotted line and the text. \textit{Right:} the same as the middle panel but for \DS data.
%\AGK{Do we want to show SELG(z) as well, on this plot? Also, the x-label is cut off on the right hand plot. Also, the text claims a good agreement on both overall shape and fine features between zeta and OIIR, but it looks from the figure that there are some fine features that don't match, for instance, the spikes at z=1.22 and 1.35 in zeta.}
%\jiaxi{fig. 1 left: the $\zeta_{\rm fac}$ curve does not really captures all the spikes in the data curves, which is the kind of the goal of the approach, no? is that problematic? maybe not; but I feel that it could be worth it to comment about that}
}
\label{fig:ssr_trend_tsnr}
\end{figure}

First, we compute the median \oii flux ratio as a function of redshift, i.e., $\oiirel$. It is an empirical measure of how difficult it is to obtain a successful ELG redshift measurement as a function of redshift, thus critical in modelling $\SSR$. We define it as 
\begin{equation}
    \oiirel = \frac{\overline{F}_{\oii}(z)\big\vert_{\tsnr<\tsnrmean}}{\overline{F}_{\oii}(z)\big\vert_{\tsnr>\tsnrmean}}\, ,
    \label{eq:oiirelative}
\end{equation}
where $\overline{F}_{\oii}$ is the median \oii flux of ELGs with reliable redshift measurements (samples from $N_{\rm goodz}$ in \cref{eq:define ssr}). $\tsnrmean\approx123$ is the median $\tsnr$ of ELGs with appropriate observing conditions (samples from $N_{\rm obs}$ in \cref{eq:define ssr}). $\oiirel$ is relevant, as the dominant factor in the redshift success criteria \cref{eq:good elg} is the SNR of the observed \oii flux. The object must have a larger \oii flux to secure a good redshift measurement when the spectrum is noisier. We, therefore, observe a larger $\oiirel$ at $z>1.5$ due to the interference of the sky emission lines to \oii measurements at that redshift range on the left panel of \cref{fig:ssr_trend_tsnr}. In DESI \Yone, we assume that \oii emitters with higher \oii flux are the same population as those with lower \oii flux, i.e., their clustering is consistent (e.g., see \cite{Gao2022} for example). $\oiirel$ is calculated at $z\in[0.8,1.6]$, the clustering measurement range, with $dz=0.01$ to have a balance between true features and noise.

%Within each redshift bin, we calculate  for good ELGs (meet \cref{eq:good elg}) with the smaller half $\tsnr$ and then divide it by that for good ELGs with the larger half of $\tsnr$. 

Next, we need to find a theoretical curve that embodies the characteristics of $\oiirel$ at $z\in[0.8,1.6]$ to build $\SSRmodel$. The theoretical squared SNR of ELG spectra $\tsnrtheory(z)$ plays a role here, defined as 
\begin{equation}
\tsnrtheory(z)=\sum_i F^2_{e,i}(z)\sigma^{-2}_{F,i}\, ,
\end{equation}
where $F_e(z)$ is the flux of emission lines\footnote{This not only includes \oii doublets, but also other lines such as \oiii, H$\beta$.} at redshift $z$ averaged over 50 processed ELG templates. 
%at $z=1$ with assumed a G-band magnitude 22\footnote{The redshift and magnitudes are not typical values. They are chosen for calculation convenience.}. 
These template spectra get their continuum removed, and their emission lines are shifted to the wavelength corresponding to redshift $z$ for $F_e(z)$ calculation. $\sigma^{-2}_F$ is the inverse variance of the sky spectra obtained in a standard exposure and averaged over the 10 DESI spectrographs\footnote{Specifically, we use exposure 165078 observed on January 28th, 2023.}. $i$ is the wavelength index, and thus, the sum is performed over all DESI wavelength for $\lambda \in [3600,9800]\,$\AA~. Thus, this quantity estimates how an ELG target's cumulative squared SNR should change with its redshift $z$. Numerically, we compute $\tsnrtheory(z)$ with a redshift resolution of $dz=0.001$ at redshift $z \in[0,1.6]$ to embody the influence of sky-residual spikes on the redshift measurement. 

We find that a simple transformation of $\tsnrtheory(z)$ as follows:
\begin{equation}
    \zeta_{\rm fac}(z) =
    \begin{cases}
    1+\frac{1500/\tsnrtheory(z)-1}{15/(z-1)},  0<z<1.6 \\
    1, \text{otherwise}\\
    \end{cases}
    ,
\end{equation}
match both the general and fine features in $\oiirel$ as shown in the left panel of \cref{fig:ssr_trend_tsnr} despite small deviations. In this way, we assign a $\zeta_{\rm fac}$ value to all ELGs with successful redshifts to represent the difficulty of obtaining such a good redshift measurement. 

%\jiaxi{from \url{https://desi.lbl.gov/trac/raw-attachment/wiki/PubBoard/PublicationReview/DESI-2023-0289_Comments/desi-sky-perexp-ivar-iron-maindark.pdf} it looks like this doesn t capture e.g. the spike at z~1.03 (I d say because of the chosen function), and also at z~1.32 (because of the expid=165078 choice); also, one sees that this curve is ~flat at 0.8<z<1.1, which explains why there s ~no correction for those bins in fig. 2; and lastly, one sees that the "size of the step" at z>1.5 varies quite a lot across those exposures}

\subsubsection{The \texorpdfstring{$\SSR$}{SSR} Model and \texorpdfstring{$\wzfail$}{wZFAIL} Weight}
\label{sec:model-SSR}
%We use the distribution of observed ELG targets as a function of $\tsnr$ as a reference, $N_{\rm full}(\tsnr)$. We compared the number of successful redshifts with $0.8<z<1.6$, for a given $\zeta$ range, $N_{\rm good}(\zeta,\tsnr)$ to this reference. 
Now we have $\zeta_{\rm fac}(z)$ and $\tsnr$ for every ELG with proper observing conditions and we expect our $\SSRmodel$ to increase with these two quantities monotonically. In practice, we split these ELGs in 20 bins of $\zeta_{\rm fac}$, each with the same number of galaxies. In each $\zeta_{\rm fac}$ bin $i$, we fit a linear relationship $k_i\tsnr+b_i$ to $\SSR$$_{,i}$ in 10 evenly spaced bins of $80<\tsnr<200$. All fits assume Poissonian errors for $\SSR$$_{,i}$, i.e., $\epsilon_{f} = \sqrt{N_{\rm goodz}}/N_{\rm obs}$.

We then take the 20 slopes $k_i$ and their corresponding median $\overline{\zeta}_{\rm fac,i}$ and perform a least-squares linear fit (errors are assumed to be the same given the number of galaxies was the same for each $k_i$ calculation) to obtain
\begin{equation}
    k = C_1\zeta_{\rm fac} + C_0.
\end{equation}
This process is repeated separately for the \DN and \DS regions, and the results, including the best-fit lines and parameters, are shown in the middle and right panel of \cref{fig:ssr_trend_tsnr}. Note that the largest 90\% $\zeta_{\rm fac}$, though having significantly larger $k_i$, agrees with the $k$-$\zeta_{\rm fac}$ trend of the smaller $\zeta_{\rm fac}$. Therefore, we do not implement more complicated regression and stability checks of these two $k$-$\zeta_{\rm fac}$ relations in this study. We will discuss the consequence of this choice in \cref{sec:ssr_impact}.
%\jiaxi{Figure 1 right panel: Have you checked the stability of the linear curve fits? It seems to me the slope strongly depends on the last two points for \DN.} 
%$C_1=0.033,C_0=-0.033$ for ELGs in the \DN region and for those in the \DS footprint, they follow $C_1=0.042,C_0=-0.043$. 
%\jiaxi{"I m wondering what is the motivation of splitting N/S; because the inputs (elg emission lines, sky ivar) does not depend on the north/south; I guess there could some differences in the (z, $\oiirel$) distribution between north/south? don t know.. but probably worth it to comment a bit} 

Finally, we convert the parameterized slope to a weight that can be attached to the galaxy catalogues to remove any trends between $\SSR$ and ($\tsnr$,\,$z$). We expect the redshift success should only increase as a function of $\tsnr$. Thus, we clip $k$ such that it has a minimum value of 0:
\begin{equation}
    k(\zeta_{\rm fac}) =
    \begin{cases}
     C_1 \zeta_{\rm fac}+C_0,  k>0 \\
    0, \text{otherwise}.\\
    \end{cases}
\end{equation}
Assuming the intercept $b=1-k\tsnrmean$, with $\tsnrmean$ being the median $\tsnr$ of all \elg targets with appropriate observing conditions, we now have a $\SSRmodel$ as a function of redshift $z$ and $\tsnr$
\begin{equation}
    \SSRmodel(\tsnr,z) = k(\zeta_{\rm fac}[z])(\tsnr-\tsnrmean)+1.
\end{equation}
We simply use the inverse of this for the weight to attach to the galaxy catalog
\begin{equation}
    \wzfail(\tsnr,z) = 1/\SSRmodel(\tsnr,z).
\end{equation}

%{\bf AJR suggest making Fig. 1 just the slope plot. Then, instead of the 0.01<z<2 plot, show the dz 0.1 plots, which validate that our crazy method at least works to null the trends with redshift. Then, something like: We note that this method, while more complicated, does a much better job of nulling the trends observed in the $\Delta z = 0.1$ redshift bins shown, compared to, e.g., using an approach similar to that applied to other DESI DR1 tracers, which model the redshift success as a function of $\tsnr$ and the fiber flux. Such an approach actually makes the results in the redshift bins that are relatively unaffected by sky lines considerably worse than neglecting to weight at all.}
\begin{figure}
\centering
\includegraphics[width=\textwidth]{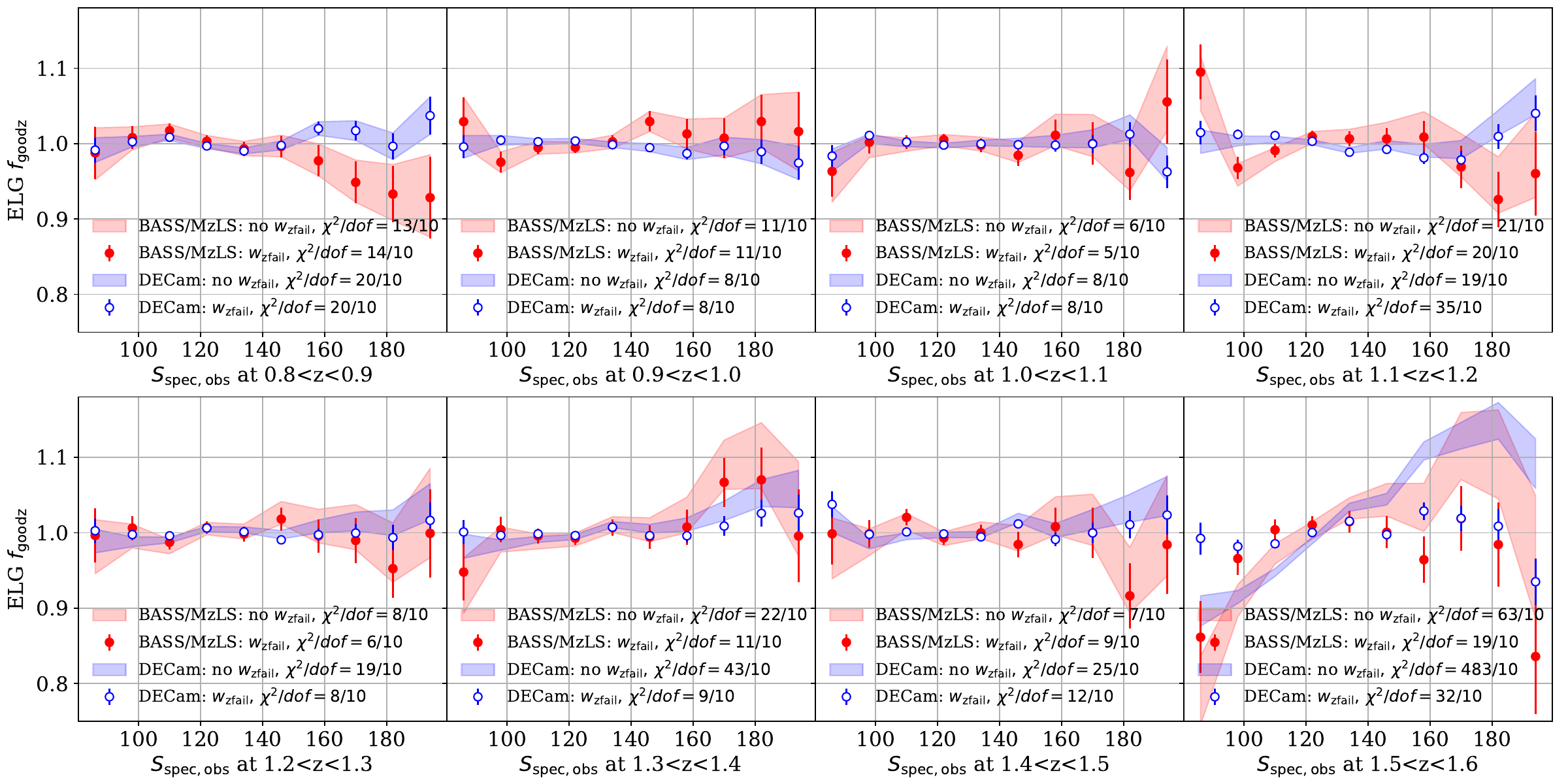}
\caption{The $\SSR$-$\tsnr$ relation before (shades) and after (error bars with dots) $\wzfail$ correction in fine redshift bins at $0.8<z<1.6$ with $dz=0.1$. \DN and \DS results are presented in red/closed circles and blue/open circles, respectively. $\wzfail$ removes $\SSR$ dependences on $\tsnr$ at $z>1.2$, especially at $1.5<z<1.6$, where the sky emissions are big issues for ELGs by construction. }
\label{fig:ssr_correction_dz0.1}
\end{figure}

\cref{fig:ssr_correction_dz0.1} presents the $\SSR$-$\tsnr$ relation before and after $\wzfail$ correction for samples from \DN and \DS surveys. 
%\jiaxi{note that you claim several times in the paper that the success rate should increase with $\tsnr$ (which is what we expect); however, that s obviously not the case in several curves in this figure; I feel this should be commented, as this sounds contradictory (but I admit I don t have on top of my head an explanation here... maybe it s just noise; but then we should say it)} 
ELGs with good redshifts are split into finer bins with $dz=0.1$ at $0.8<z<1.6$ for $\SSR$ computation. The errors are binomial errors $\epsilon_{\rm zfail}$ 
\begin{equation}
    \epsilon_{\rm zfail} = \frac{\sqrt{N_{\rm obs}(1-N_{\rm goodz}/N_{\rm obs})}}{N_{\rm obs}}
\end{equation}
and their $\chi_{\rm zfail}^2=(\SSR-1)^2/\epsilon_{\rm zfail}^2$ represents the deviation from a uniform $\SSR$-$\tsnr$ relation. The mean values of $\SSR$ fluctuate a bit, but all relations are consistent with monotonic tendencies. The $\SSR$-$\tsnr$ dependency in both footprints is close to unity before the $\wzfail$ correction, except for the $\sim10\%$-level variations at $1.3<z<1.4$ and the $\sim$30\%-trend at $1.5<z<1.6$. They correspond to the big spikes of $\oiirel$ and $\zeta_{\rm fac}$ in the left panel of \cref{fig:ssr_trend_tsnr}. Therefore, $\wzfail$ based on $\zeta_{\rm fac}$ suppress the dependences of $\SSR$ on $\tsnr$ at $1.2<z<1.6$, especially on $1.5<z<1.6$ given the decreasing $\chi_{\rm zfail}^2$. Note that the $\chi_{\rm zfail}^2/\rm dof$ values can still be as large as 3. But we do not worry about it as they represent a sub-percent $\SSR$ difference, leading to a minor impact on the clustering as presented in \cref{sec:ssr_impact}. 
%\AGK{I think we should interpret the total chi2 we get after applying weights. This is 93.5 for N and 131.1 for S, if I added them correctly! There are 80 data points, and I believe 2 free parameters. So the N results are consistent with no trend, while the S results are not as good; however, we can argue that additional corrections have negligible impact on clustering.}\jiaxi{Could you remind me where you get those two numbers? I added the TSNR2 relation after $\wzfail$ correction but the $\chi^2$ are different from what you cite here}\AGK{I probably made an arithmetic mistake.. I'm trying to add the numbers associated with the points in Fig.\ 2.}

\subsection{The Focal Plane Correction \texorpdfstring{$\czfail$}{etafail}} 
\label{sec:ssrcorr}
%\jiaxi{Do the authors have any intuition as to why fgoodz is dependent on distance from focal plane center? The source of this dependency is not immediately apparent to me}

%\jiaxi{Is it clear that the focal plane and $\wzfail$ weights are separable?}

\begin{figure}
\centering
\includegraphics[width=0.9\textwidth]{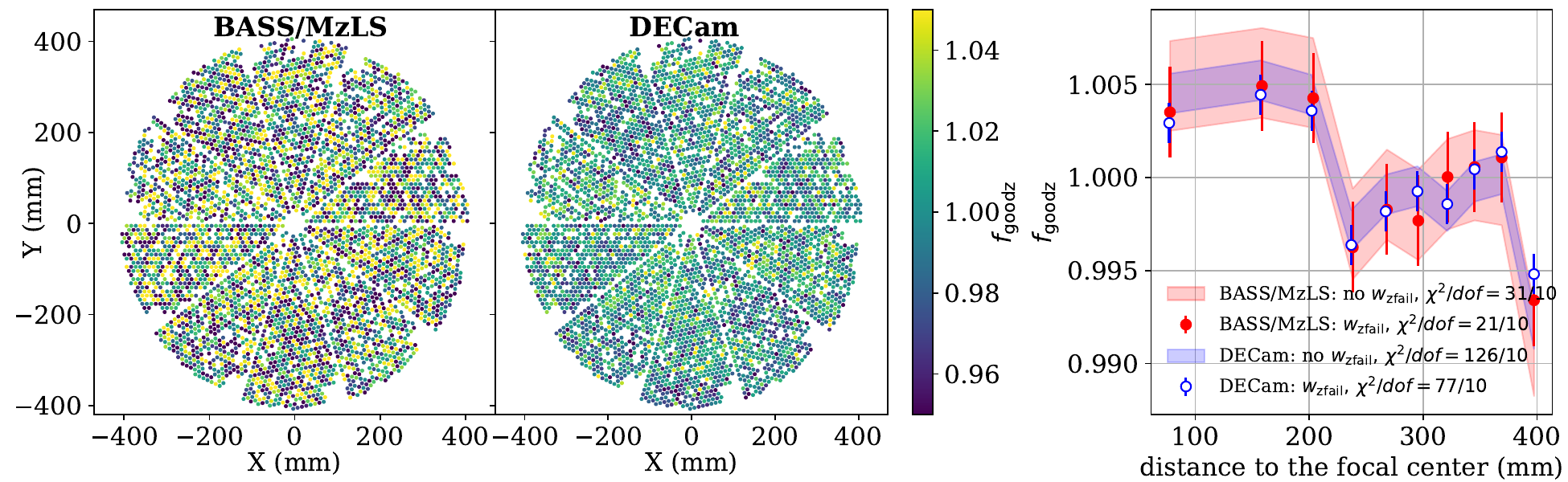}
\caption{\textit{Left:} The $\SSR$ (with $\wzfail$) of each fibre distributed on the focal plane for \DN survey. The $\SSR$ differences on the focal plane can be as large as 10\%. \textit{Middle:} the same as the left panel but for \DS survey. Its $\SSR$ variations are smaller than that of \DN survey. \textit{Right:} $\SSR$ as a function of the distance to the focal centre without (shades) and with (error bars) $\wzfail$ corrections. Red/closed circles and blue/open circles represents results from \DN and \DS respectively. The $\SSR$ variations on the focal plane are independent of $(\tsnr,\,z)$, thus $\wzfail$ does not eliminate this dependence. }
\label{fig:ssr_trend_focal}
\end{figure}

The $\SSR$ not only depends on $(\tsnr,\,z)$. It also varies across the focal plane (e.g., \cite{eBOSS_ELG_2021}). But $\tsnr$ is defined for each exposure, without information on the focal plane. The definition of $\zeta_{\rm fac}$ used quantities that are averaged over spectrographs, eliminating the information as well. Consequently, $\wzfail$ does not remove the $\SSR$ variations on the focal plane as shown in \cref{fig:ssr_trend_focal}. The left and middle panel shows a 10\% $\SSR$ variation across the focal plane, resulting in $\chi_{\rm zfail}^2/{\rm dof}=4753/4262$ for \DN and $\chi_{\rm zfail}^2/{\rm dof}=4716/4262$ for \DS where 4262 is the number of fibres. Note that the median number of good redshift measurements per fibre for \DN survey is 101, and that for \DS is 575. The smaller statistics of \DN lead to larger $\SSR$ variations and error bars, thus similar $\chi_{\rm zfail}^2$ to \DS. 
%These correspond to $\chi_{\rm zfail}^2/{\rm dof}=4761.5/4262$ for \DN and $\chi_{\rm zfail}^2/{\rm dof}=5283.6/4262$ for \DS where 4262 is the number of fibres. Moreover, $\wzfail$ does not greatly improve the $\SSR$ distribution on the focal plane,  introducing $\Delta \chi_{\rm zfail}^2=+88.1$ for \DN and $\Delta \chi_{\rm zfail}^2=-297.1$ for \DS. 
Marginalising over the angular direction, the $\SSR$ of fibres that are close to the focal centre shows 1\% higher $\SSR$ than that of the distant fibres as shown in the right panel of \cref{fig:ssr_trend_focal}. This difference still exists with $\wzfail$ corrections despite the fact that it is already close to the ideal case of a uniform distribution. Therefore, we need an extra correction factor on $\wzfail$ as a function of its position on the focal plane (or a fibre-dependent correction). 

%As introduced in \cref{sec:wzfail}, $\wzfail$ significantly reduces the dependency of $\SSR$ on $\tsnr$, but it is imperfect. There is a 1\% $\SSR$ difference between small and large $\tsnr$ after applying $\wzfail$. In addition to this, \cref{fig:ssr_trend_focal} presents 10\%-level variations in the $\SSR$ on the focal plane after $\wzfail$. This corresponds to $\chi_{\rm \DN}^2/{\rm dof}=4849.6/4262$ and $\chi_{\rm \DS}^2/{\rm dof}=4986.5/4262$. The $\SSR$ averaged on the radial direction shows a 1\% $\SSR$ difference in the focal plane's centre and outskirts (see \cite{fiberoverview_claire} for detailed explanations from the fibre-system side). Moreover, $\wzfail$ does not greatly improve the $\SSR$ distribution on the focal plane, only introducing $\Delta \chi_{\rm \DN}^2=-88.1$ and $\Delta \chi_{\rm \DS}^2=-297.1$. Therefore, a correction $\eta$ based on the 2D distribution of the $\SSR$ over the focal plane is probably better than another correction with $\tsnr$.

The correction of $\wzfail$ is defined fiber-wise as:
\begin{equation}
    \eta_{\rm zfail,i} = \overline{f}_{\rm goodz}(i)/{f}_{\rm goodz}(i),
    \label{eq:wzfail_correction}
\end{equation}
where $i$ is the fibre ID ranging from 0 to 4999, but only 4262 of them provide $\SSR$ measurements. The other 738 fibres are not used for scientific observations\footnote{They mostly are stuck fibres, but we can still use them to obtain sky spectra}. ${f}_{\rm goodz}(i)$ is the $\SSR$ of fibre $i$ and $\overline{f}_{\rm goodz}(i)$ is the mean $\SSR$ of all samples. $\czfail$ null all $\SSR$ on the focal plane by definition, and thus, the $\SSR$ dependence on the distance to the focal centre is removed in both the \DN and \DS footprints. This fibre-wise correction is then implemented on each observed ELG target, and thus, the new total weight $w'_{\rm tot}$ is written as
%multiplied to $w_{\rm tot}$ when we activate this weight. 
\begin{equation}
w'_{\rm tot} = w_{\rm comp}w_{\rm sys} (\wzfail \czfail),
\label{eq:new wtot}
\end{equation}
where $\wzfail \czfail$ is the corrected redshift failure weight.

\begin{figure}
\centering
%\includegraphics[scale=0.3]{preliminary_figures/SSR_model_ZFAIL_TSNR2_ELG_v1.2.png}
%\caption{The $\SSR$ dependencies on $\tsnr$, normalized throughput in g (\texttt{fibf5z}), r (\texttt{fibf6}) and z (\texttt{fibf8}) bands without (lines with shades) and with $\wzfail$ (error bars). The remaining $\SSR$ trends not corrected by $\wzfail$ were removed by $\wzfail\czfail$.
\includegraphics[width=\textwidth]{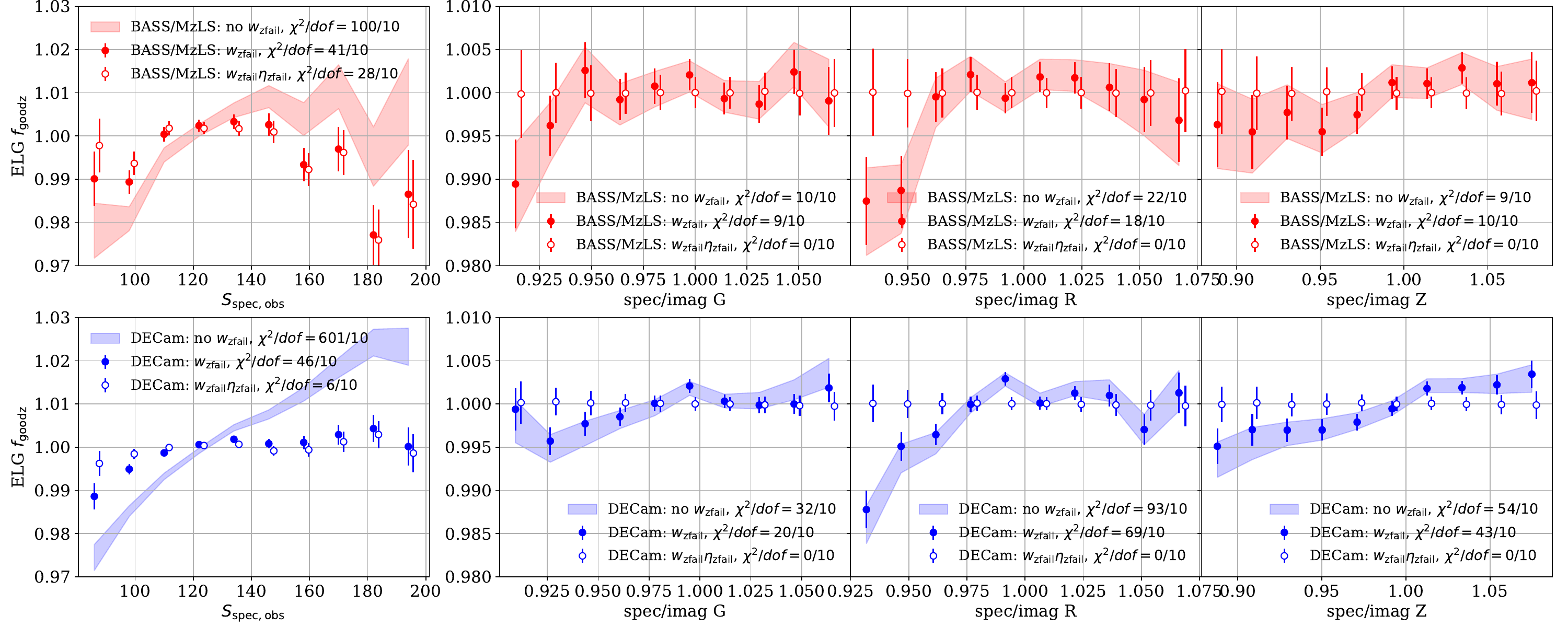}
\caption{The $\SSR$ dependencies on $\tsnr$ (the first column) and on averaged ratios of spectroscopic and imaging flux of standard stars in G (the second column), R (the third column) and Z (the fourth column) bands without $\wzfail$ (lines with shades), with $\wzfail$ (closed error bars) and with $\wzfail\czfail$ (open error bars) for \DN regions (the upper panel) and \DS regions (the lower panel). 
The remaining $\SSR$ trends not corrected by $\wzfail$ were removed by $\wzfail\czfail$.
}
\label{fig:ssr_wzfail}
\end{figure}

\subsection{Performances of \texorpdfstring{$\wzfail$}{wzfail} and \texorpdfstring{$\czfail$}{etafail}}
\label{sec:ssr_impact}
We have proved in \cref{sec:wzfail} and \cref{sec:ssrcorr} that $\wzfail$ and $\czfail$ can respectively correct the $\SSR$ dependences on $\tsnr$ in fine redshift bins and the focal plane by construction. In this section, we will check their performance of removing other $\SSR$ trends and their influences on galaxy clustering. 

In the first column of \cref{fig:ssr_wzfail}, the dependence of $\SSR$ on $\tsnr$ for ELGs targets without redshift selection is around 5\% (shades) and $\wzfail$ suppresses the trend down to 1\% (closed error bars). The improvements are also significant, as shown in the decreasing $\chi_{\rm zfail}^2$ values in both \DN and \DS regions. Therefore, $\wzfail$ is included in the ELG LSS catalogue as part of the correction of observational systematics (\cref{eq:wtot}). 
%\AGK{I think we should state, very clearly, that we choose to apply wzfail in the catalogs, and not etazfail; and we should give the reason for this choice. }
The over-suppression of \DN $\SSR$ for those with large $\tsnr$ ($\sim \tsnrtheory$) probably come from the imperfect linear regression on small $\zeta_{\rm fac}$ values as shown in the middle panel of \cref{fig:ssr_trend_tsnr}. The open circles in the same subplots illustrate the effects of $\czfail\wzfail$ on the same $\SSR$ trends as in \cref{fig:ssr_wzfail}. This extra correction brings back missing information on the focal plane by $\wzfail$ and, therefore, improves $\SSR$ weighting in small $\tsnr$. This agrees with our conclusion that the source of the over-correction of $\SSR$ in \DN area is from the $\wzfail$ modelling, not from any uncorrected $\SSR$ trend on the focal plane. We also notice that $\zeta_{\rm fac}$ is not a perfect representation of $\oiirel$. We will develop a better $\SSRmodel$ for future data release as these aspects a minor effect on the clustering. 

The second to fourth panels of \cref{fig:ssr_wzfail} illustrates the $\SSR$ dependence on the spectrograph-to-imaging flux ratio of standard stars at $\lambda\in [4500,5500]\,$\AA~(G band), $\lambda\in[6000,7300]\,$\AA~(R band) and $\lambda\in[8500,9800]\,$\AA~(Z band). They represent the fibre aperture loss in the central wavelength of these three bands \cite{Spectro.Pipeline.Guy.2023}. As the definition of $\tsnr$ includes the throughput of the fibre (\cref{eq:tsnr}), we expect that $\SSR$ varies with throughputs, and $\wzfail$ can also remove most of the $\SSR$ dependences on throughputs. The fibre aperture loss in R and Z bands present a more significant trend than in G band as the \oii emissions of DESI \elg targets are observed in R and Z bands. But the $\SSR$ dependences on the throughputs remain after applying $\wzfail$ as shown in \cref{fig:ssr_wzfail}. This is consistent with our finding in \cref{fig:ssr_trend_focal} that $\wzfail$ by construction does not include the modelling of $\SSR$ variations on the focal plane. The new weight, $\czfail\wzfail$, nullifies the $\SSR$ dependences on the throughputs by construction. This is because $\czfail$, defined per-fibre, eliminates the 2D $\SSR$ variations on the focal plane. Consequently, any quantities defined based on fibres will have a unity mean $\SSR$ after applying $\czfail$. However, such an \textit{ad hoc} correction might have the risk of overfitting the \Yone, so it is not part of the LSS weight budget. We may consider adding it to $w_{\rm tot}$ in the future data release.
%\AGK{(I would be concerned about overfitting in the definition of etazfail, potentially removing real small-scale clustering).}
%\AGK{The overall trends in the N region are still bad, right? Deltachisq = 16, which is about 4-sigma bad using the sqrt(2*N) rule.}

\begin{figure}
\centering
\includegraphics[width=\textwidth]{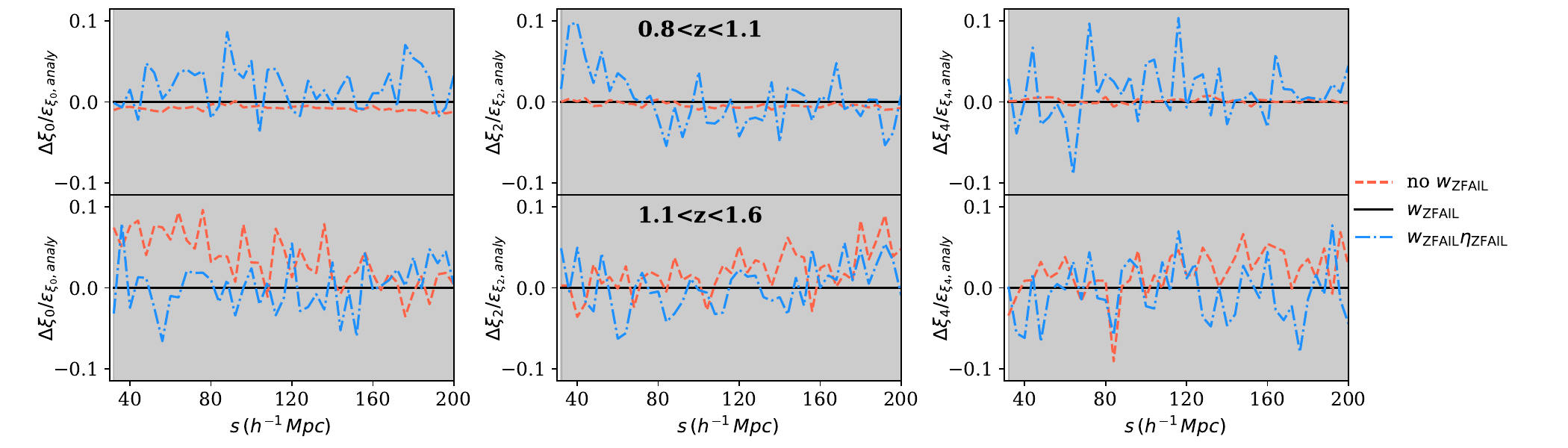}
\caption{The comparison of 2PCF monopole (first column), quadrupole (second column) and hexadecapole (third column) for ELGs implementing the standard weighting scheme $w_{\rm tot}$ (`$\wzfail$', solid lines), the $w_{\rm tot}$ but excluding $\wzfail$ (`no $\wzfail$', dashed lines) and the new total weight $w'_{\rm tot}$ (`$\wzfail\czfail$', dash-dotted lines) at $20<s<200\mpch$. The first row is for ELGs at $0.8<z<1.1$, and the second is for $1.1<z<1.6$. The clustering influences of redshift failure weights are minor, with a difference of $<0.05\varepsilon_\xi$.}
\label{fig:wzfail_effect_mps}
\end{figure}

\begin{figure}
\centering
\includegraphics[width=\textwidth]{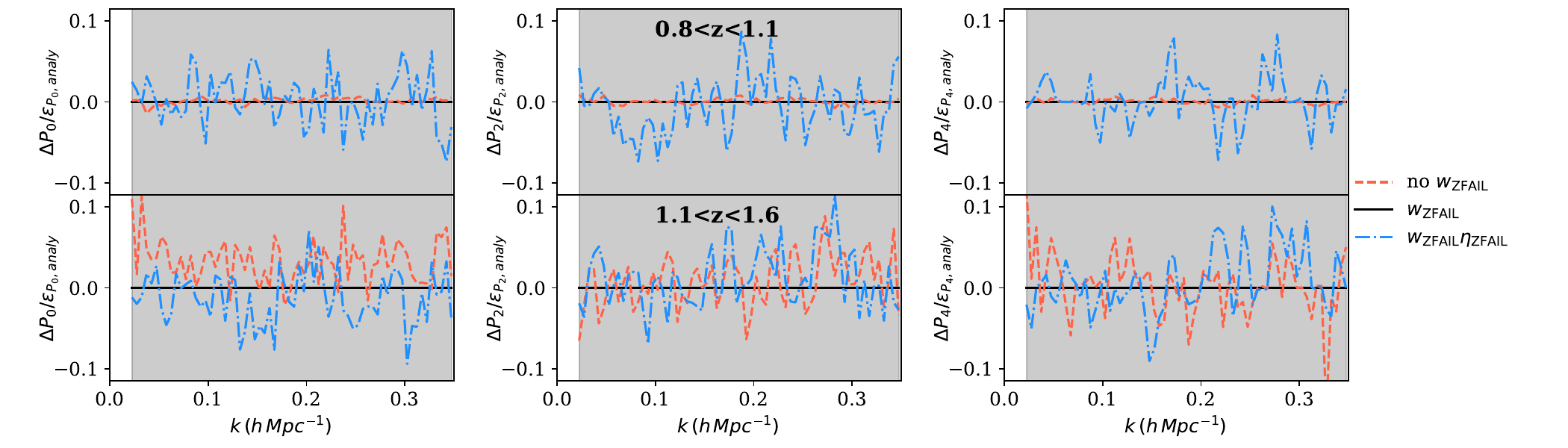}
\caption{Similar to \cref{fig:wzfail_effect_mps} but for power spectra multipoles $P_\ell(k)$ at $0.02<k<0.35\invmpch$. The clustering influences of redshift failure weights in the Fourier space are as small as those in the configuration space.}
\label{fig:wzfail_effect_pk}
\end{figure}

We then study the effects of $\wzfail$ and $\wzfail\czfail$ on galaxy clustering. To do this, we calculate the 2PCF $\xi_\ell(s)$ and power spectra $P_\ell(k)$ of ELGs from the LSS catalogues at $0.8<z<1.1$ and $1.1<z<1.6$ (redshift bins for LSS analyses). The galaxy weight of ELGs is the product of $w_{\rm FKP}$ and the total observational weight in three formats: $w_{\rm tot}$ (\cref{eq:wtot}), $w_{\rm tot}/\wzfail$ (no corrections on the failed redshift measurements) and $w'_{\rm tot}$ (\cref{eq:new wtot}, the complete LSS weights with the new redshift failure weight). They are denoted as `$\wzfail$', `no $\wzfail$' and `$\wzfail\czfail$' respectively in \cref{fig:wzfail_effect_mps} and \cref{fig:wzfail_effect_pk}. These two figures present the clustering effects of the redshift failure weights compared to the standard $w_{\rm tot}$ weight in the configuration space and Fourier space. Both weights have $<0.05\varepsilon$ influence on all two-point statistics multipoles at both redshift bins (see \cref{sec:chi2-comparison} for the definition of the 1-$\sigma$ clustering error $\varepsilon$). Such a minor effect of the redshift failure weights has been observed in eBOSS ELGs as well \cite{eBOSS_ELG_2021}. $\wzfail$ leads to a larger clustering impact at $1.1<z<1.6$ compared to the clustering at $0.8<z<1.1$. This is implied in \cref{fig:ssr_correction_dz0.1} as the significant corrections that $\wzfail$ accomplished are all at $1.1<z<1.6$. The effect of $\wzfail\czfail$ on redshift ranges and multipoles are all comparable, showing the impact of the 2D $\czfail$ correction is evenly applied to all samples. 

\begin{table}
\centering
{\renewcommand{\arraystretch}{1.3}
\begin{tabular}{|c|c|c|c|c|c|}
\hline
Space & redshift & {$\chi^2_{{\rm sys},\,\text{no }\wzfail}$} &   {$\chi^2_{{\rm sys},\,\wzfail\czfail}$} \\ \hline
\multirow{2}{*}{Config.}
& $0.8<z<1.1$ & 0  &0.2  \\ 
& $1.1<z<1.6$ & 0.15 & 0.18  \\ \hline
\multirow{2}{*}{Fourier} 
& $0.8<z<1.1$ & 0& 0.22 \\
& $1.1<z<1.6$ &0.32&  0.25 \\
\hline
\end{tabular}
\caption{The $\chi_{\rm sys}^2$ values for the ELG clustering calculated with different weights compared to the standard clustering with the $w_{\rm tot}$ weight in both configuration space and Fourier space for $\ell=0,2,4$. The third column shows ELGs implementing $w_{\rm tot}$ but excluding $\wzfail$, and the fourth column is for ELGs applying $w'_{\rm tot}$ (\cref{eq:new wtot}), which includes the corrected failure weight $\wzfail\czfail$. In both configuration and Fourier spaces, the influences of the redshift failure weight and its correction are minor regardless of the redshift range.}
\label{tab:SSR_chi2}
} 
\end{table}

\cref{tab:SSR_chi2} provides the $\chi_{\rm sys}^2$ values of ELG clustering computed without $\wzfail$ and with the new total weight $w'_{\rm tot}$ (\cref{eq:new wtot}) compared to the standard clustering computed with $w_{\rm tot}$. These values are no larger than 0.32. So, we conclude that the cosmological impact of $\wzfail$ and its correction $\czfail$ can be neglected. Nevertheless, it is important to note that $\SSR$ not only depends on the effective exposure time $\tsnr$ and redshift $z$. $\SSR$ of each fibre also varies on the focal plane, independent of the variations brought by ($\tsnr$,\,$z$). Therefore, future surveys must check the dependence of $\SSR$ on different observing conditions and their correlation to develop a complete weight to correct the failed redshift measurements.

\section{Redshift Catastrophics and Redshift Uncertainty} 
\label{sec:catas}
Catastrophics and redshift uncertainty are systematics hidden in \elg samples with secure redshift measurements (\cref{eq:good elg}). With $\Delta v = \Delta zc/(1+z)$ from repeated observations (\cref{sec:repeats}), the catastrophics refer to pairs of redshift with a large $|\Delta v|$. 
There are several definitions of `a large $|\Delta v|$', which are $|\Delta v| >1000\kms$ \cite{Dawson2015,ELG.TS.Raichoor.2023} (our definition), $\Delta v>5\sqrt{\langle \Delta v^2}\rangle$ \cite{sys2001_2df}, or those with clear evidence of misidentification of emission lines or between the emission line and the sky residuals \cite{sys2010blunder}. 
All these definitions mean that the radial position of a small number of tracers is shifted to a large extent, leading to imprints on the clustering. ELGs from DESI EDR and SDSS-IV eBOSS have $f_{\rm catas}\sim0.3\%$ \cite{eBOSS_ELG_2021,ELG.TS.Raichoor.2023}. Given such a small fraction, its clustering and cosmological impacts are neglected in eBOSS analyses. 
%Its cosmological impact was corrected by a factor of (1-$f_{\rm catas})^{-2}$ in the theoretical clustering in \cite{sys2010blunder} with a catastrophics rate $f_{\rm catas}\approx5\%$. However, \cite{sys2016_fastsound} found this correction might be inaccurate. 
%Moreover, the catastrophics of line emitters (i.e., ELGs) are poorly understood due to the limited statistics of good spectra. 
%The well-resolved \oii-doublet of ELG spectra and a large amount of repeated observation in DESI enable in-depth research on this topic. 

Redshift uncertainty comes from redshift measurement and thus exists in every galaxy in LSS catalogues. Repeated observations without the catastrophics samples provide a statistical estimation via the width of the $\Delta v$ distribution (e.g. \cite{QSOdv2018,LRGdv2020,ELG.TS.Raichoor.2023} from SDSS-IV eBOSS). Note that repeated observations cannot capture the velocity shift of QSOs, which is part of QSO redshift uncertainty. But we can study this type of mixed uncertainty via their clustering effects (see \cite{RedrockQSO.Brodzeller.2023,mine}). For galaxy samples, we assume statistical uncertainty represents the redshift uncertainty. The redshift uncertainty of DESI ELGs from EDR is $\sim 8\kms$, leading to little impact on the clustering \cite{ELG.TS.Raichoor.2023,mine}. 

Given their properties, neither catastrophics nor redshift uncertainty can be corrected target-wise like $\wzfail$ but they influence the clustering collectively. Catastrophics of line emitters (i.e., ELGs) are poorly understood due to the limited statistics and small fraction as mentioned above. Now, with millions of reliable ELG samples from DESI \Yone \cite{DESI2024.I.DR1} and well-modelled, realistic ELG mocks \cite{antoine2023}, we have enough good-quality samples to study the effect of catastrophics carefully. DESI \Yone have fewer samples with long exposure compared to EDR \cite{DESI2023b.KP1.EDR} to guarantee better redshift measurements. Therefore, we will revisit these two aspects in DESI \Yone ELGs, understand the data, model them with galaxy mocks, and see how to prevent them from biasing the cosmological measurements.

\begin{figure}[htb]
\centering
\includegraphics[width=0.8\textwidth]{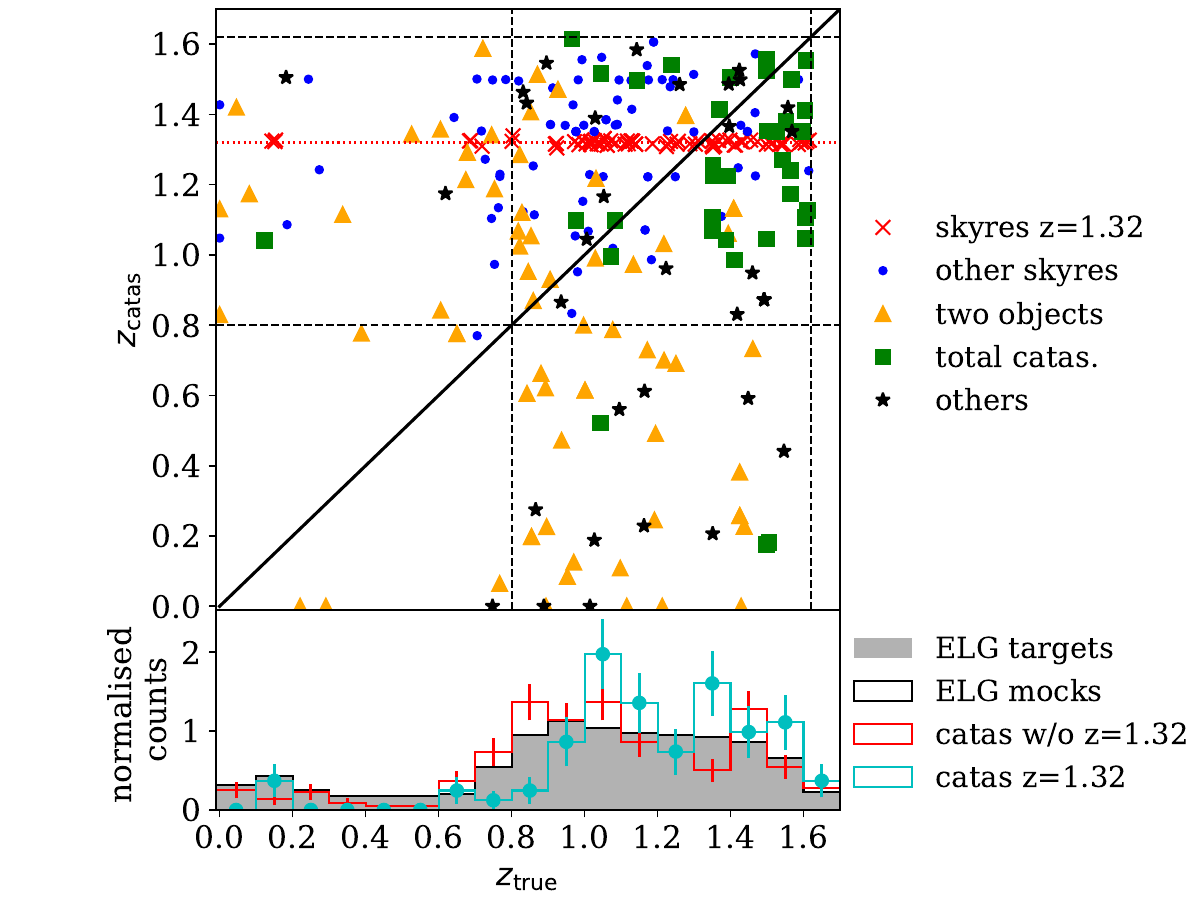}
\caption{\textit{Top:} The comparison between the \textit{true} redshift and the \textit{catastrophics} redshift for catastrophics of DESI ELGs. The exceptions are spectra that contain two objects (triangles), which means that both redshifts are true redshifts, and spectra that have no emission lines for redshift determination (squares), which means that both redshifts are catastrophically failed redshifts. The cross, small dots represent the redshift misidentification of the sky residuals to be the \oii emission that concentrates on $z_{\rm catas}\approx1.32$ (the horizontal dotted line), the same misidentification of sky residuals at other wavelengths. Stars are all other types of catastrophics. The dashed lines in $z=0.8$ and $z=1.6$ indicate the redshift range of ELGs for clustering measurements. \textit{Bottom:} The (true) redshift distribution of the \full catalogue (grey shades), the ELG \abacus mocks (solid line), the catastrophics without $z_{\rm catas}\approx1.32$ (step histogram with error bars only), and the catastrophics with $z_{\rm catas}\approx1.32$ (step histogram with dotted error bars). The error bars of both catastrophics samples are assumed to be binomial errors. The redshift distribution of two types of catastrophics roughly follows the \full catalogue. 
}
\label{fig:catas_z1-z2}
\end{figure}
%\textcolor{red}{what kind of ELGs has the risk of being misidentified? (TS paper)}
\subsection{Characterising the Catastrophics}
\label{sec:catas_feature}
Catastrophics are discrepancies in the redshift measurements for the same object. As reported in \cref{sec:repeats}, 0.26\% of the repeated ELG pairs have $|\Delta v|>1000\kms$. There is no preference for their locations in the focal plane, and their redshift distributions are shown in \cref{fig:catas_z1-z2}. The upper panel of \cref{fig:catas_z1-z2} compares the \textit{true} redshift ($z_{\rm true}$) and the \textit{catastrophics} redshift ($z_{\rm catas}$) for the ELG catastrophics detected with the repeated observation catalogue. We determine the $z_{\rm true}$ and $z_{\rm catas}$ via visual inspection of their reduced spectra with the help of \texttt{prospect}\footnote{\url{https://github.com/biprateep/prospect}} \cite{VIGalaxies.Lan.2023,VIQSO.Alexander.2023}. 
If a redshift measurement matches secure spectral features (e.g., the \oii doublets or multiple emission lines), this redshift is $z_{\rm true}$ and the other measurement on the same object is $z_{\rm catas}$. There are five types of catastrophics for ELGs shown in \cref{fig:catas_z1-z2} (see \cref{appendix:catas-spec} for examples of their spectra):
\begin{itemize} 
    \item The misidentification of the residuals of sky emissions at around 8600--8700$\,$\AA~ to be the \oii emission (sky confusion hereafter). They form a prominent feature in \cref{fig:catas_z1-z2} at $z_{\rm catas}\approx1.32$ as shown in crosses, comprising 26.7\% of the total catastrophics samples. We observe a significant doublet-like spike in one of their spectra at 8600--8700\AA, which is difficult to smooth out and, therefore, is identified as the \oii emission. The evolving DESI redshift pipeline might resolve this issue by improving the sky subtraction and line-identification process \cite{Spectro.Pipeline.Guy.2023}.
    %\jiaxi{are you sure of that? is it not just due to variable strength of the sky emission lines in this 8600-8700 A range? ie some sky-subtraction issue? also, the Bailey+ ref may not be relevant here, as this paper is for redrock (ie the redshift fitting), this issue rather sounds like a desispec issue} 
    There is no such feature in the other spectrum of these repetitively observed objects, and thus, the redshift obtained by the other spectrum is correct ($z_{\rm true}$).
    %In our sample, $z_{\rm sky}\approx 1.32,\,1.35,\, 1.5$ as shown in \cref{fig:catas_z1-z2}, constitutes $11.5\%,\,2.2\%,\,1.3\%$ of the total failures; 
    \item The misidentification of the sky residuals at other wavelengths as the \oii emission. The sky emissions/residuals appear at $\lambda>8000\,$\AA~, corresponding to $z_{\rm catas}>1.1$. As the \oii flux of ELGs at $z_{\rm true}<1.1$ is systematically lower than the sky residuals (see Figure 15 of \cite{ELG.TS.Raichoor.2023}), this type of catastrophics tends to increase the redshift of ELGs, i.e., from $z_{\rm true}<1.1$ to $z_{\rm catas}>1.1$. They make up 27.6\% of the total catastrophic pairs. 
    \item Two objects in the spectra. The fibre might capture two overlapping objects with different redshifts, or the ELG spectra were contaminated by a bright, nearby object ($z_{\rm catas}$ or $z_{\rm true}$ around 0). As there are two sets of spectral features and thus true redshifts, the redshift pipeline can take either as its output redshift. These samples are mainly at $0.6<z<0.9$ and comprise 22.0\% of the total catastrophics. 
    \item Total catastrophics. There are 12.1\% catastrophics that do not have a correct redshift measurement at all. This is because the \oii emission is too faint to be found (by the \textsc{Redrock} pipeline and visual inspection). Therefore, \textsc{Redrock} can identify any of sky residuals as \oii doublets. Their $z_{\rm true}$ and $z_{\rm catas}$ concentrate at $z>1.1$ because most of the sky residuals are at $\lambda>8000\,$\AA~ as mentioned above.
    \item Other catastrophics. The remaining 11.6\% catastrophics include line confusion between \oii doublets and other emission lines such as \oiii and H$\,\alpha$, bad spectra, and QSO spectra misclassified as ELGs. 
    %\item The confusion between the emission lines (line confusion hereafter). e.g., the \oii doublet is misidentified as the \oiii line, and vice versa. \jiaxi{Its true redshift obtained by visual inspection is usually xx. } This pattern is not significant compared to the random catastrophics in our sample \jiaxi{as indicated by the cyan lines ($z_2=\lambda_{\oiii}/\lambda_{\oii}(1+z_1)-1$ for the solid line and $z_2=\lambda_{\oii}/\lambda_{\oiii}(1+z_1)-1$ for the dashed line) in \cref{fig:catas_z1-z2} (a)}. 
\end{itemize}

Although there are five patterns of catastrophics, we only need to model the excess sky confusion and all the other patterns (random catastrophics hereafter). The random catastrophics alone cannot reproduce the distinct feature of $z_{\rm catas}\approx 1.32$. Random catastrophics with additional sky confusions should represent more general catastrophics while avoiding overfitting our specific sample. 

\begin{figure}[htb] 
\centering
\includegraphics[width=0.8\textwidth]{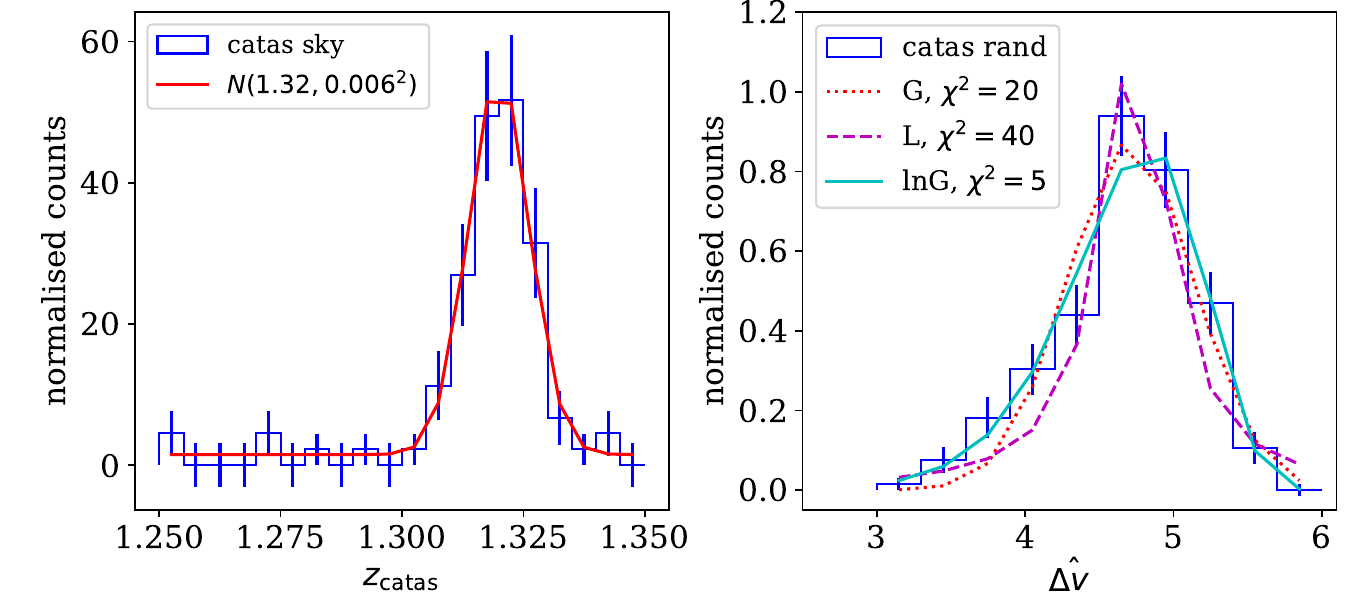}
\caption{\textit{Left:} The zoom-in histogram of $z_{\rm catas}$ of the sky confusion samples at $1.25<z<1.35$ (empty histogram). The solid line is the best-fitting Gaussian function $\mathcal{N}(1.32,0.006^2)$ with an offset of 0.61. They compose of 24.9\% of catastrophics. \textit{Right:} The histogram of $\hat{\Delta v}$ of the random catastrophics (empty histogram with binomial errors). It is fitted by a Gaussian profile (`G', the dotted line), a Lorentzian profile (`L', the dashed line), and a log-normal profile (`lnG', the solid line). The $\chi^2$ improvement brought by the extra 1 parameter of the log-normal distribution is significant. Therefore, $\mathcal{LN}(0.64,0.25^2,6.62)$ is the best description of the random catastrophics $\hat{\Delta v}$. }
\label{fig:catas_hist}
\end{figure}

The $z_{\rm catas}$ distribution of the sky confusion (left) and the $\text{log}_{10}(|\Delta v|)$ ($\hat{\Delta v}$ hereafter) distribution of random catastrophics (right) are presented in \cref{fig:catas_hist}. We fit the excess sky confusion with a Gaussian distribution $\mathcal{N}(\mu_{\rm sky},\sigma_{\rm sky}^2)$ where $\mu_{\rm sky} =1.32$ and $\sigma_{\rm sky}=0.006$, enclosing 24.9\% of the total catastrophics. This Gaussian function is shifted up by 0.61, representing the remaining 1.9\% sky confusion that has been included in random catastrophics. 
%\AGK{I don't understand what you mean by this. A Gaussian is characterized by only two params, mean and variance}
The $\hat{\Delta v}$ distribution of the random catastrophics can be fitted with a Gaussian profile, a Lorentzian profile and a log-normal profile with an extra free parameter $\mathcal{LN}(\mu_{\rm ran},\sigma_{\rm ran}^2, \hat{v_0};\hat{\Delta v})$ which is defined as
\begin{equation}
    \mathcal{LN}(\mu_{\rm ran},\sigma_{\rm ran}^2, \hat{v_0};\hat{\Delta v}) = \frac{A}{\sqrt{2\pi}\sigma_{\rm ran}(-\hat{\Delta v}+\hat{v_0})}\exp\Bigl\{-\frac{\left[\ln(-\hat{\Delta v}+\hat{v_0})-\mu_{\rm ran}\right]^2}{2\sigma_{\rm ran}^2}\Bigr\}.
    \label{eq:lorentzian}
\end{equation}
Assuming a binomial error, the log-normal profile provides a much better fitting to the $\hat{\Delta v}$ distribution of random catastrophics. Its best-fitting parameters are $A= 0.99,\; \mu_{\rm ran}=0.64, \; \sigma_{\rm ran}=0.25, \;  \hat{v_0}=6.62$. 

\subsection{Modelling the Catastrophics}
\label{sec:catas_modelling}
The $z_{\rm true}$ of sky confusion and random catastrophics both follow the $n(z)$ of the ELG targets (\full catalogue) as illustrated in \cref{fig:catas_z1-z2}. Therefore, we randomly take 0.26\% galaxies from the \abacus mocks and implement the following pattern to reproduce the realistic catastrophics 
\begin{equation}
    z_{\rm catas}=
    \begin{cases}
        z_{\rm true}+\Delta v_{\rm ran}, &  \text{random catastrophics, 76\%}; \\
        \mathcal{N}(1.32,0.006^2), &  \text{extra sky confusion, 24\%}, \\
    \end{cases}
    \label{eq:catas-apply}
\end{equation}
where $z_{\rm true}$ is the redshift of \abacus galaxy mocks. Half of $\Delta v_{\rm ran}$ is positive and the other half is negative, and $|\Delta v_{\rm ran}|$ follows $\mathcal{LN}(0.64,0.25^2,6.62)$. We also realise a 1\% random catastrophics 
%and 0.1\% sky confusion 
on \abacus mocks for demonstration purposes. 1\% is an upper limit of the ELG catastrophics rate in eBOSS and DESI \citep{Dawson2015,ELG.TS.Raichoor.2023}. In addition, we remove \abacus ELGs at $1.31<z<1.33$, which is the easiest implementation on data if we want to avoid the impact on the sky confusion at $z\approx1.32$. 
%\jiaxi{0.1\% sky residuals produce a similar KL divergence \AR{explain KL} to the current \full catalogue in $n(z)$ distribution at $1.3<z<1.34$.} 
Given a similar $n_{\rm local}(z)$ after implementing the catastrophics, the $w_{\rm FKP}$ of samples with $z_{\rm catas}$ is obtained with the $n_{\rm local}(z_{\rm catas})$. All the patterns are implemented on 25 realizations of \abacus galaxy mocks.

\begin{figure}[htb]
\centering
\includegraphics[width=\textwidth]{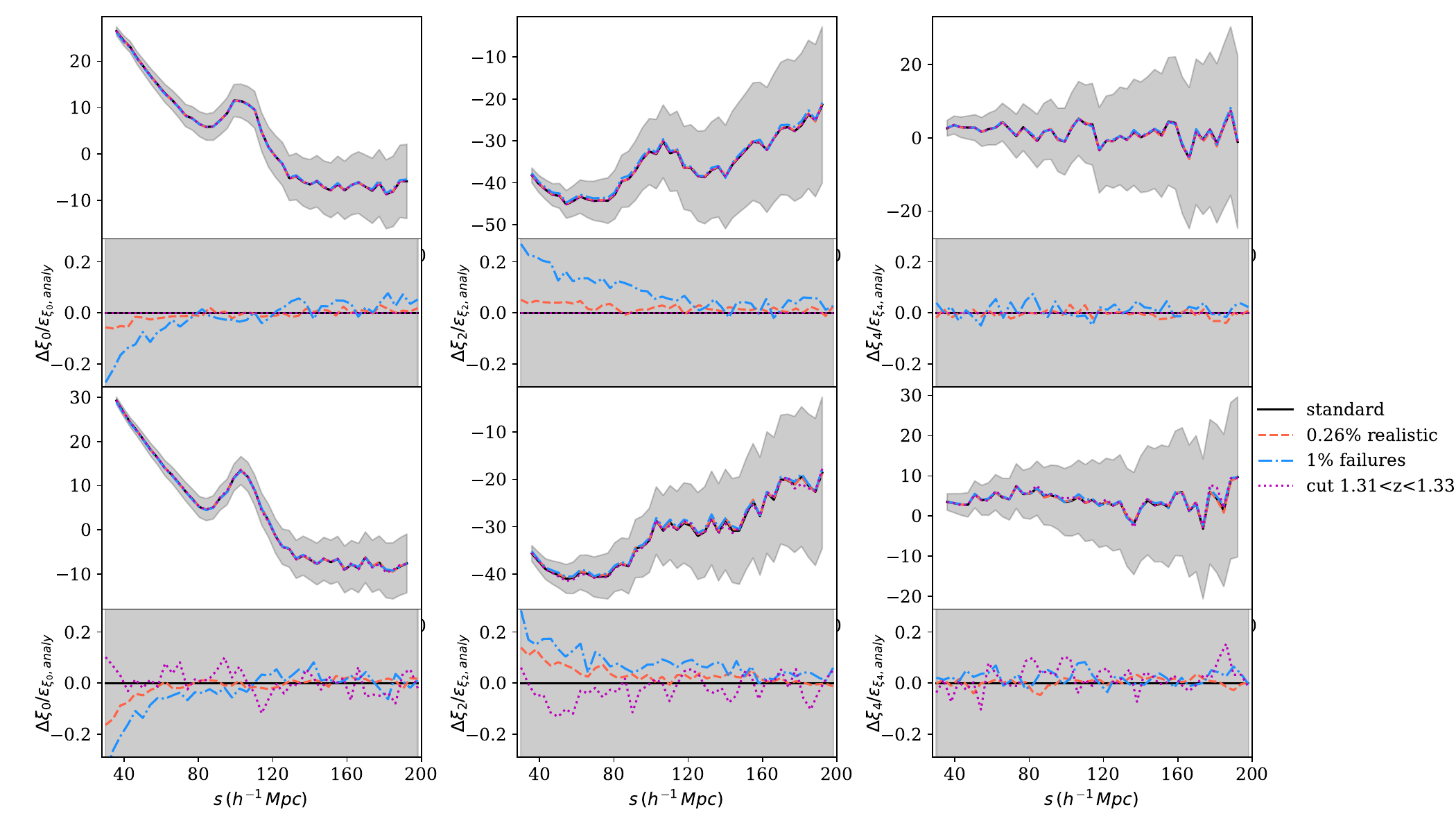}
\caption{The 2PCF monopole (first column), quadrupole (second column) and hexadecalpole (third column) of model galaxies from the standard \abacus mocks (solid lines), \abacus mocks with 0.26\% realistic catastrophics (dashed lines), \abacus mocks with hypothetical 1\% random catastrophics (dash-dotted lines) and \abacus mocks without $1.31<z<1.33$ (dotted lines). The grey areas are the 1-$\varepsilon_\xi$ errors of clustering obtained via analytical covariances (\Cref{sec:abacusmocks}). The first and the second rows are the 2PCF multipoles and the differences between the standard \abacus mocks and catastrophics/cuts rescaled by the errors provided by the analytical covariance matrices at $0.8<z<1.1$. The third and fourth rows are similar but model galaxies at $1.1<z<1.6$. Catastrophics lead to negligible clustering effects at $0.8<z<1.1$ except for the hypothetical 1\% case. For galaxies at $1.1<z<1.6$, the realistic 0.26\% catastrophics shift the monopole and quadrupole by $>0.1\varepsilon_\xi$ at $s<60\mpch$, similar to the effects of the hypothetical catastrophics. Hexadecapoles are robust to the catastrophics effect. }
\label{fig:catas_clustering_mps}
\end{figure}

\begin{figure}[htb]
\centering
\includegraphics[width=\textwidth]{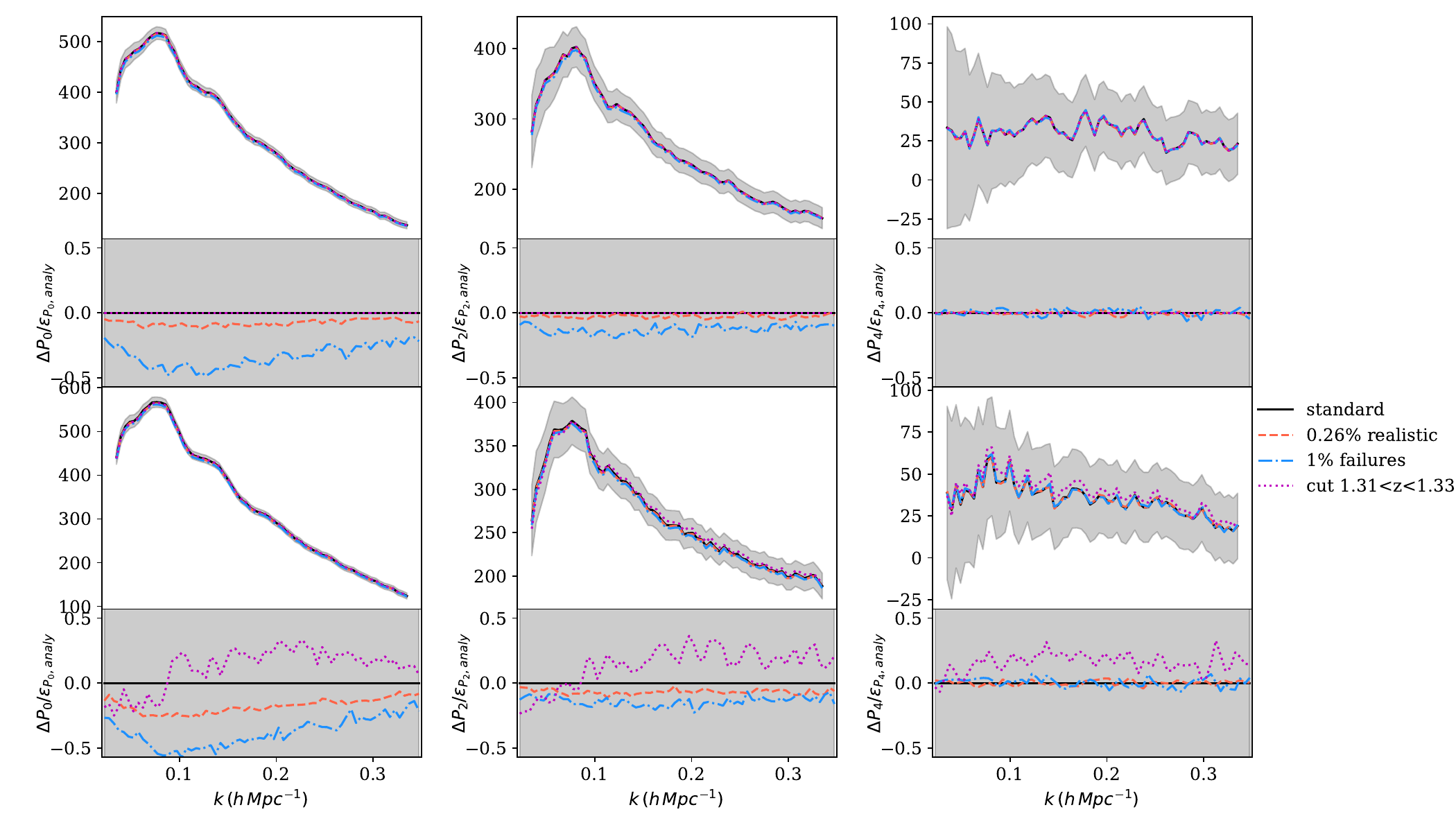}
\caption{Similar to \cref{fig:catas_clustering_mps} but for power spectra $P_\ell(k)$ $\ell=0,2,4$.}
\label{fig:catas_clustering_pk}
\end{figure}

\subsection{Impact of Catastrophics}
\label{sec:catas_cosmo}
We present in \cref{fig:catas_clustering_mps} and \cref{fig:catas_clustering_pk} the impact of realistic 0.26\% catastrophics, hypothetical 1\% random catastrophics and removing ELGs at $1.31<z<1.33$ (the sky-confusion-contaminated range) on multipoles of the 2PCF $\xi_\ell$ and power spectra $P_\ell$. Removing ELGs at $1.31<z<1.33$ is a possible solution to avoid the impact of the hidden sky confusion at $z\approx1.32$, corresponding to $f_{\rm catas}=2.54\%$. 

Their influences are mainly embodied in galaxies at higher redshift $1.1<z<1.6$. Realistic and 1\% random catastrophics result in up to $0.2\varepsilon_\xi$ impact on $\xi_0$ and $\xi_2$ on $s<60\mpch$. The catastrophics can also suppress the power spectra monopole by up to $0.5\varepsilon_P$. %This highlights how destructive sky confusion can be. This is because 0.06\% of such failures (25\% of the realistic failures) can result in equivalent impacts to 0.8\% of the random catastrophics (hypothetical random catastrophics minus realistic random catastrophics). 
Removing model ELGs at $1.31<z<1.33$ slightly perturbs the 2PCF as data and randoms both implemented the truncation. So there is no bias in its 2PCF and the difference from the standard clustering is due to the reduction of galaxies. Removing ELGs boosts the amplitude of ELG $P_\ell$ at $1.1<z<1.6$ by up to $0.3\varepsilon_P$ while the other catastrophics suppress the clustering. %The opposite trend to the catastrophics models might be caused by the difference in the window functions. 

%\AGK{It seems a bit odd that the 1\% and 0.26\% catastrophics rate are similar to each other, and both quite different from 0\% catastrophic rate. I would have expected more of a continuous change.}

At $0.8<z<1.1$, only 1\% random catastrophics lead to up to $0.2\varepsilon_\xi$ and $0.5\varepsilon_P$ influence in the configuration space and Fourier space, respectively. This is because the number of galaxies that the realistic catastrophics interferes at $0.8<z<1.1$ is too small, and removing galaxies at $1.31<z<1.33$ does not affect the LSS at lower redshifts. Hexadecapoles are robust to all three implementations. 

Their $\chi_{\rm sys}^2$ are presented in \cref{tab:catas_chi2}. The DESI BAO analysis focusing on the configuration space will not \cite{DESI2024.III.KP4} be influenced by any of the listed catastrophics. This is because the catastrophics do not move the position of the BAO peak. Their maximum $\chi_{\rm sys}^2$ (the left column of each $\chi_{\rm sys}^2$) is as small as 0.32. Both lead to negligible changes in the BAO cosmological parameters. However, their impacts become larger in Fourier space at $1.1<z<1.6$ due to the systematical shift they cause in the monopole and quadrupole. This could interfere with the RSD measurements. We should remind our reader that all catastrophics and redshift-removal implemented to \abacus ELG mocks result in the change of window function. The difference in $P_\ell$ and $\chi_{sys}^2$ values is not equivalent to their influence on cosmological measurements. Therefore, we perform cosmological fitting with their corresponding window function to determine how they will interfere with the parameters. 
\begin{table}
\centering
{\renewcommand{\arraystretch}{1.3}
\begin{tabular}{|c|c|c|c|c|c|c|c|}
\hline
Space & redshift & {$\chi^2_{\rm sys,\,realistic}$} &  {$\chi^2_{\rm sys,\,random}$} & {$\chi^2_{\rm sys,\,cut}$} \\ \hline
\multirow{2}{*}{Config.}
& $0.8<z<1.1$ & 0.05 &0.33  & /  \\
& $1.1<z<1.6$ &  0.1 &  0.38 &  0.42 \\ \hline
\multirow{2}{*}{Fourier} 
& $0.8<z<1.1$ &  0.25 & 4.39 & /  \\
& $1.1<z<1.6$ &1.46& 6.66 &4.21 \\
\hline
\end{tabular}
\caption{Similar to \cref{tab:SSR_chi2} but for catastrophics models: 0.26\% realistic failures, 1\% random catastrophics, removing ELGs at $1.31<z<1.33$ compared to the \abacus standard clustering. The catastrophics all present larger impacts for ELGs at $1.1<z<1.6$. In configuration space, their influence is minor. However, in Fourier space, the catastrophics might bias the cosmological measurement. }
\label{tab:catas_chi2}
} 
\end{table}

\begin{figure}[htb]
\centering
\includegraphics[width=\textwidth]{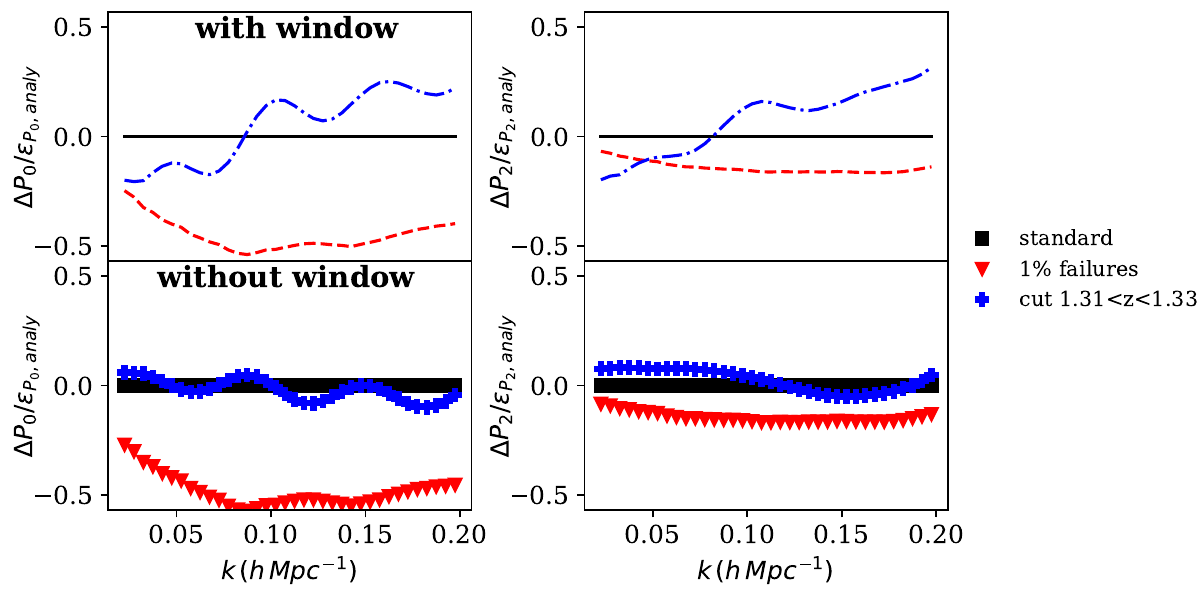}
\caption{The best-fit theoretical $P(k)$ at $1.1<z<1.6$ for the standard clustering (solid lines or squares), the clustering with hypothetical 1\% failures (dashed lines or flipped triangles) and the clustering with truncation at $1.31<z<1.33$ (dash-dotted lines or plus signs). The first column is the $P(k)$ monopole, and the second is the quadrupole. The upper panel shows the best-fit theoretical curves with window functions applied (i.e., agree with mock $P(k)$), while the lower panel shows the same curves without window functions. The $P(k)$ differences for 1\% failures did not shift much without the window function, while that of the cut $1.31<z<1.33$ changed its values. This implies that the nuisance parameters absorb the clustering effect of the 1\% failures, whereas the truncation effect is primarily accounted for by its window function.}
\label{fig:catas_windows_pk}
\end{figure}

The differences in the best-fitting values of cosmological parameters between the catastrophics clustering and the standard clustering in Fourier space are shown in \cref{tab:catas_cosmo_fullmodelling} and \cref{tab:catas_cosmo_shapefit}.  The maximum-likelihood values $\chi^2$ /dof (\cref{eq:chi2_cosmo}) of all these samples are all smaller than $3.2/59$ as the clustering is calculated from mocks based on accurate $N$-body simulations and thus is fitted well by theories. Despite the large $\chi^2_{\rm sys}$, the cosmological impacts of all types of systematics are no larger than 0.2$\sigma$. That is because the nuisance parameters and window functions absorb most of the effects. The nuisance parameters adjust the shape and amplitude of the theoretical curves directly, while the window function influences $P(k)$ through geometric effects. As shown in \cref{fig:catas_windows_pk}, the clustering difference between the 1\% failures and the standard case remains largely unchanged with or without the window function. This suggests that the catastrophics parameters absorb the nuisance effects rather than the window function for the catastrophics-contaminated mocks. In contrast, the clustering difference between the truncated and standard samples varies when the geometric effects (window function) are ignored. This indicates that the clustering changes caused by truncating a slice of the redshift result from changes in geometry. To conclude, the influence of catastrophics effects on DESI ELG in DR1 is not a concern.

\begin{table}
\centering
{\renewcommand{\arraystretch}{1.3}
\begin{tabular}{|c|c|c|c|c|c|c|c|}
\hline
Type & redshift & {$\sigma_h$} &  {$\sigma_{\omega_{\rm cdm}}$} & {$\sigma_{\omega_b}$}& {$\sigma_{\As}$} \\ \hline
\multirow{2}{*}{Realistic}
& $0.8<z<1.1$ & 0 &0  &0 &0 \\
& $1.1<z<1.6$ & 0.04 &  0.01 &  0 & 0 \\ \hline
\multirow{2}{*}{Random} 
& $0.8<z<1.1$ & 0 & 0 &0 &0.06  \\
& $1.1<z<1.6$ & 0.04 & 0.01 &0 &0.08 \\ \hline
\multirow{2}{*}{Cut} 
& $0.8<z<1.1$ &  / & / & / & /  \\
& $1.1<z<1.6$ & 0.12 & 0.05 &0 &0.17\\
\hline
\end{tabular}
\caption{The cosmological impact of catastrophics in full modelling measurements: 0.26\% realistic failures, 1\% random catastrophics, removing ELGs at $1.31<z<1.33$ compared to the \abacus standard clustering. $\sigma$ represent the 68\% confidence level of DESI \Yone cosmological constraints. All influences are smaller than 0.2$\sigma$. }
\label{tab:catas_cosmo_fullmodelling}
} 
\end{table}

\begin{table}
\centering
{\renewcommand{\arraystretch}{1.3}
\begin{tabular}{|c|c|c|c|c|c|c|c|}
\hline
Type & redshift & {$\sigma_{\alpha_{\rm iso}}$} &  {$\sigma_{\alpha_{\rm AP}}$} & {$\sigma_{df}$}& {$\sigma_{dm}$} \\ \hline
\multirow{2}{*}{Realistic}
& $0.8<z<1.1$ & 0.04 &0.03  &0.01 &0.01 \\
& $1.1<z<1.6$ & 0 &  0.02 &  0 & 0.03 \\ \hline
\multirow{2}{*}{Random} 
& $0.8<z<1.1$ &  0.04 & 0 &0.01 &0.1  \\
& $1.1<z<1.6$ & 0.06 & 0.02 &0 &0.07 \\ \hline
\multirow{2}{*}{Cut} 
& $0.8<z<1.1$ &  / & / & / & /  \\
& $1.1<z<1.6$ & 0.06 & 0.04 &0.04 &0.14\\
\hline
\end{tabular}
\caption{Similar to \cref{tab:catas_cosmo_fullmodelling} but for ShapeFit results. The cosmological impacts are all smaller than 0.2$\sigma$.}
\label{tab:catas_cosmo_shapefit}
} 
\end{table}
%\AGK{These largely look like broadband changes, but do we want to forward-reference analyses looking at the impact of these weights on fnl as well as RSD?}

%\jiaxi{\cite{sys2010blunder} propose a $(1-f_{\rm catas})^2$ correction on the theory to take into account catastrophics. We check this factor by comparing it with the clustering ratio between the \abacus mocks with catastrophics and the standard \abacus mocks. As demonstrated in \cref{fig:catas_ratio_mps} and \cref{fig:catas_ratio_pk}, $(1-f_{\rm catas})^2$ is not a good description of the impact of clustering of catastrophics. This is shown in the study of \cite{sys2010blunder,sys2016_fastsound}. \jiaxi{In 5.3, mention the ratioanle behind the $(1-f)^2$ factor - it should work for non-clustering catastrophics also in the randoms. Are the catastrophics clustered? }}

\subsection{Redshift Uncertainty} 
\label{sec:zerr}
\begin{figure}[htb]
\centering
\includegraphics[scale=0.55]{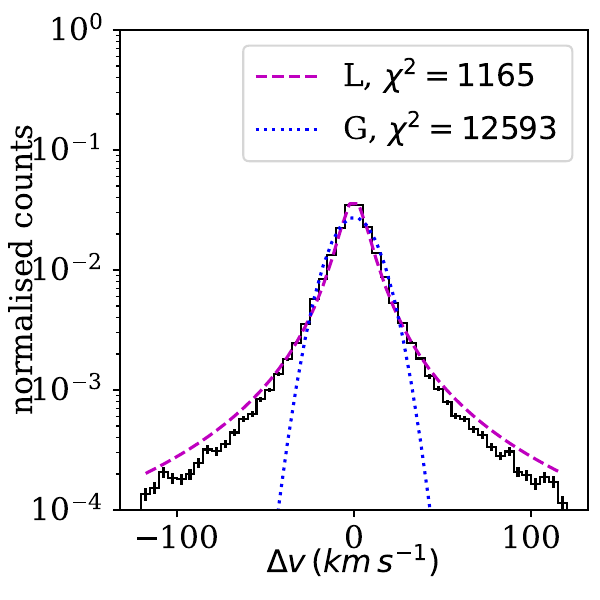}
\caption{The distribution of DR1 ELG redshift difference $\Delta v$ (histogram with binomial error bars) that are smaller than 1000$\kms$. There are the best-fit Gaussian profile $\mathcal{N}(0.06,12.7^2)$ (`G', the dotted line) and Lorentzian profile $\mathcal{L}(0.03,8.5)$ (`L', the dashed line). The Lorentzian profile is a better description of the ELG redshift uncertainty than a Gaussian profile.}
\label{fig:redshift_uncertainty}
\end{figure}
%\jiaxi{"this is smaller than the EDR ELG uncertainty" (8.82 km/s vs. 10.7 km/s): why is that? they re supposed to be the same ELGs, with a ~similar pipeline (iron vs. fuji), etc. if anything, the EDR has longer effective times (1200s I think); and from mostly EDR data, Raichoor+23 and Lan+23 found ~8 km/s=> did you use in Yu+24 the ELG\_LOP\_notQSO sample in 0.8<zreliable<1.6? (note that in sv3/elg\_hip = main/elg\_lop...)}
Repeated observations at $0.8<z<1.6$ with $|\Delta v|<1000\kms$ are used to measure the redshift uncertainty of ELGs as shown in \cref{fig:redshift_uncertainty}. Compared to a Gaussian profile, a Lorentzian profile described as follows is a better model of \Yone ELG $\Delta v$ distribution
\begin{equation}
    \mathcal{L}(p,w_{\Delta v}) = \frac{A}{1+((x-p)/w_{\Delta v})^2}.
    \label{eq:dv_redshift_uncertainty}
\end{equation}
where $A = 0.04$, $p=0.03$ and $w_{\Delta v}=8.5\kms$. This $w_{\Delta v}$ is similar to the values of the visual inspection and survey validation programs for DESI ELGs \cite{VIGalaxies.Lan.2023,ELG.TS.Raichoor.2023} that are at the level of $\sim 10\kms$, thus its impact on the \Yone ELG clustering is negligible on $s>5\mpch$ \cite{mine}. Its percentile velocities at 50, 95, and 99.5 \% are 8.2, 53.5 and 139.0$\kms$, respectively. They are less than half of the values from eBOSS ELGs, which are 20, 100, and 300$\kms$ \cite{eBOSS_ELG_2021} thanks to the improvement of spectrographs and redshift pipeline \cite{DESI2022.KP1.Instr,Spectro.Pipeline.Guy.2023,Redrock.Bailey.2024}. 

\section{Conclusion} 
\label{sec:conclusion}
In this paper, we investigate the spectroscopic systematics of ELGs in DESI \Yone and their clustering impact on 2PCF $\xi(s)$ and the power spectrum $P(k)$. Spectroscopic systematics arise from the failed redshift measurements and cause non-cosmological variations in the galaxy distribution. One of the aspects is the redshift success rate $\SSR$ dependences on the observing conditions. The other ones are collective effects introduced by the catastrophics and the redshift uncertainty in the redshift measurement. 

The actual $\SSR$ depends on the squared signal-to-noise for ELG spectra \texttt{TSNR2\_ELG} ($\tsnr$) and redshift $z$, leading to larger $\SSR$ for ELGs with larger $\tsnr$ at different redshift bins. This is equivalent to unfairly up-weighting ELGs with high $\tsnr$. %The characteristics of $\SSR$ differ between the \DN and \DS footprints due to differences in photometry. 
We develop the redshift failure weight $\wzfail$ as the inverse of a linear function of $\tsnr$ as a function of ($\tsnr,\,z$). With $\wzfail$, we successfully recover a close-to-unity $\SSR$-$\tsnr$ relation for ELGs at fine redshift bins with $dz=0.1$. However, this weight is imperfect since there is no information for $\SSR$ variations on the focal plane. Therefore, the $\SSR$ dependencies on the distance from the focal centre, and the fibre aperture loss at G, R, and Z bands, do not show improvement after applying $\wzfail$ . We thus create fibre-based corrections $\czfail$ on the $\wzfail$. $\czfail$ is the inverse of the $\wzfail$-weighted $\SSR$ for each fibre. Implementing $\czfail$ leads to a more uniform $\SSR$-$\tsnr$ relation and a unity relation between $\SSR$, the distance to the focal plane, and all the throughput in different bands. Due to the small $\SSR$ difference before any corrections, $\wzfail$ and $\wzfail\czfail$ result in smaller than $0.05\varepsilon_\xi$ in the monopole, quadrupole, and hexadecapole of $\xi$ and $P(k)$, corresponding to less than $\chi^2_{\rm sys}=0.25$ differences. It means that their cosmological influences are negligible. 

Catastrophics and redshift uncertainty are systematics that are difficult to subtract from individual targets. They can be explored through repeated observations with the redshift difference $\Delta v$, and their clustering impact can be modelled in cosmological theories. The catastrophics rate of DESI \Yone ELGs is 0.26\%, similar to that of ELGs from DESI survey validation and eBOSS. There are five types of failure patterns: sky confusions that result in $z_{\rm catas}\approx1.32$ (26.7\%), other misidentification between sky-emission residuals and \oii doublets (27.6\%), double objects (22\%), double failures (12.1\%), and other failures (11.6\%). 
%Among them, the repeated observations for double objects are correct and those of double failures are wrong. Thus, the single-measurement failure rate is not simply half of the failure rate measured by repeated observation. 

Despite the variety of the patterns, we can model their $z_{\rm catas}$-$z_{\rm true}$ relation in the \abacus galaxy mocks with 75.1\% of random catastrophics and 24.9\% of the excess sky confusions besides the random catastrophics (realistic catastrophics). We also generate catastrophics-contaminated galaxy mocks with a hypothetical with 1\% random catastrophics and mocks without the sky-confusion-contaminated redshift range $1.31<z<1.33$. %The clustering effect of the 0.26\% realistic catastrophics and the 1\% random catastrophics is similar. This suggests that sky confusion can impact LSS more than random catastrophics. 
Their impacts on the 2PCF are as small as $0.3\varepsilon_\xi$ for all multipoles at $0.8<z<1.6$, corresponding to $\chi_{\rm sys}^2<1$. Therefore, we do not expect catastrophics to affect the BAO and RSD measurement in the configuration space. However, realistic failures and 1\% random catastrophics systematically suppress the power spectra by up to $0.5\varepsilon_P$ at both redshift bins. Removing the contaminated shell of $1.31<z<1.33$ from the data is a simple solution to avoid bias in cosmological measurements based on the configuration space. In the Fourier space, removing a small shell of the observed 3D map boosts the monopole by $0.3\varepsilon_P$ at $1.1<z<1.6$. All these lead to $\chi_{\rm sys}^2$ values larger than 1 in Fourier space. We calculate the new window function of all three mocks and do RSD fitting with full-modelling and ShapeFit compression. We find that the cosmological influence of the realistic catastrophics, the hypothetical 1\% catastrophics and removing the contaminated $1.31<z<1.33$ are smaller than 0.2$\sigma$. This means that the deviation from the standard $P_\ell$ is mainly from the change of geometry and nuisance parameters. This level of systematics is also within the systematics budget of \cite{DESI2024.V.KP5}.
%However, 
%Therefore, we leave the discussion to \cite{DESI2024.V.KP5}. In addition to these tests, we revisit the theoretical corrections of the catastrophics $(1-f_{\rm catas})^2$ proposed by \cite{sys2010blunder}. The ratio between catastrophics-contaminated clustering and standard clustering follows neither $(1-f_{\rm catas})$ nor $(1-f_{\rm catas})^2$. We need further investigation for this failure if $f_{\rm catas}$ is not negligible. The redshift uncertainty of DESI \Yone ELGs at $0.8<z<1.6$ is $w_{\Delta v}=8.82\kms$, smaller than that of the eBOSS ELGs and DESI EDR ELGs. This also means that it has a minor clustering impact at all scales.

In conclusion, the spectroscopic systematics of ELGs have a minor impact on DESI cosmological measurements. Nevertheless, any unexpected patterns in this type of systematics are still worth examining. The modelling of $\SSR$ needs to take into account the information on the focal plane as well despite the small variations. The catastrophics, composed of 0.26\% of the ELGs, have minor impacts on the cosmological parameters. This is not problematic for the current redshift surveys, but it can be a problem for photometric surveys from space. The redshift uncertainty of ELGs is small thanks to the characteristic \oii doublet. But that of the Lyman-Alpha Emitters (LAEs), as a type of emission-line galaxies, though, might be more complicated. With more data and improved telescopes, we will be able to resolve all these issues before they become a significant problem. 

%We need to develop a better correction in cosmological theory, as the current one is insufficient. The redshift uncertainty of the \oii emitters are small even up to $z=1.6$. 

%But it is not necessarily true for line emitters like Lyman-Alpha Emitters (LAEs) at higher redshift.

\section*{Data Availability}
The data used in this analysis will be made public along the Data Release 1 (details in \url{https://data.desi.lbl.gov/doc/releases/}). Zenodo includes all data to reproduce the figures in this paper: \url{https://doi.org/10.5281/zenodo.11302697}.

\begin{acknowledgments}
JY, DFS, and JPK acknowledge the support from the SNF 200020\_175751 and 200020\_207379 ``Cosmology with 3D Maps of the Universe" research grant. We would like to thank Anand Raichoor, Allyson Brodzeller, Ruiyang Zhao and Julien Guy for their helpful discussions. We also would like to thank Andrei Variu for his support in visualising DESI spectra.

This material is based upon work supported by the U.S. Department of Energy (DOE), Office of Science, Office of High-Energy Physics, under Contract No. DE–AC02–05CH11231, and by the National Energy Research Scientific Computing Center, a DOE Office of Science User Facility under the same contract. Additional support for DESI was provided by the U.S. National Science Foundation (NSF), Division of Astronomical Sciences under Contract No. AST-0950945 to the NSF’s National Optical-Infrared Astronomy Research Laboratory; the Science and Technology Facilities Council of the United Kingdom; the Gordon and Betty Moore Foundation; the Heising-Simons Foundation; the French Alternative Energies and Atomic Energy Commission (CEA); the National Council of Humanities, Science and Technology of Mexico (CONAHCYT); the Ministry of Science and Innovation of Spain (MICINN), and by the DESI Member Institutions: \url{https://www.desi.lbl.gov/collaborating-institutions}. Any opinions, findings, and conclusions or recommendations expressed in this material are those of the author(s) and do not necessarily reflect the views of the U. S. National Science Foundation, the U. S. Department of Energy, or any of the listed funding agencies.

The authors are honored to be permitted to conduct scientific research on Iolkam Du’ag (Kitt Peak), a mountain with particular significance to the Tohono O’odham Nation.

\end{acknowledgments}



\appendix

\section{Visualise Spectra of ELG Catastrophics} 
\label{appendix:catas-spec}
\begin{figure}
\centering
\includegraphics[width=\textwidth]{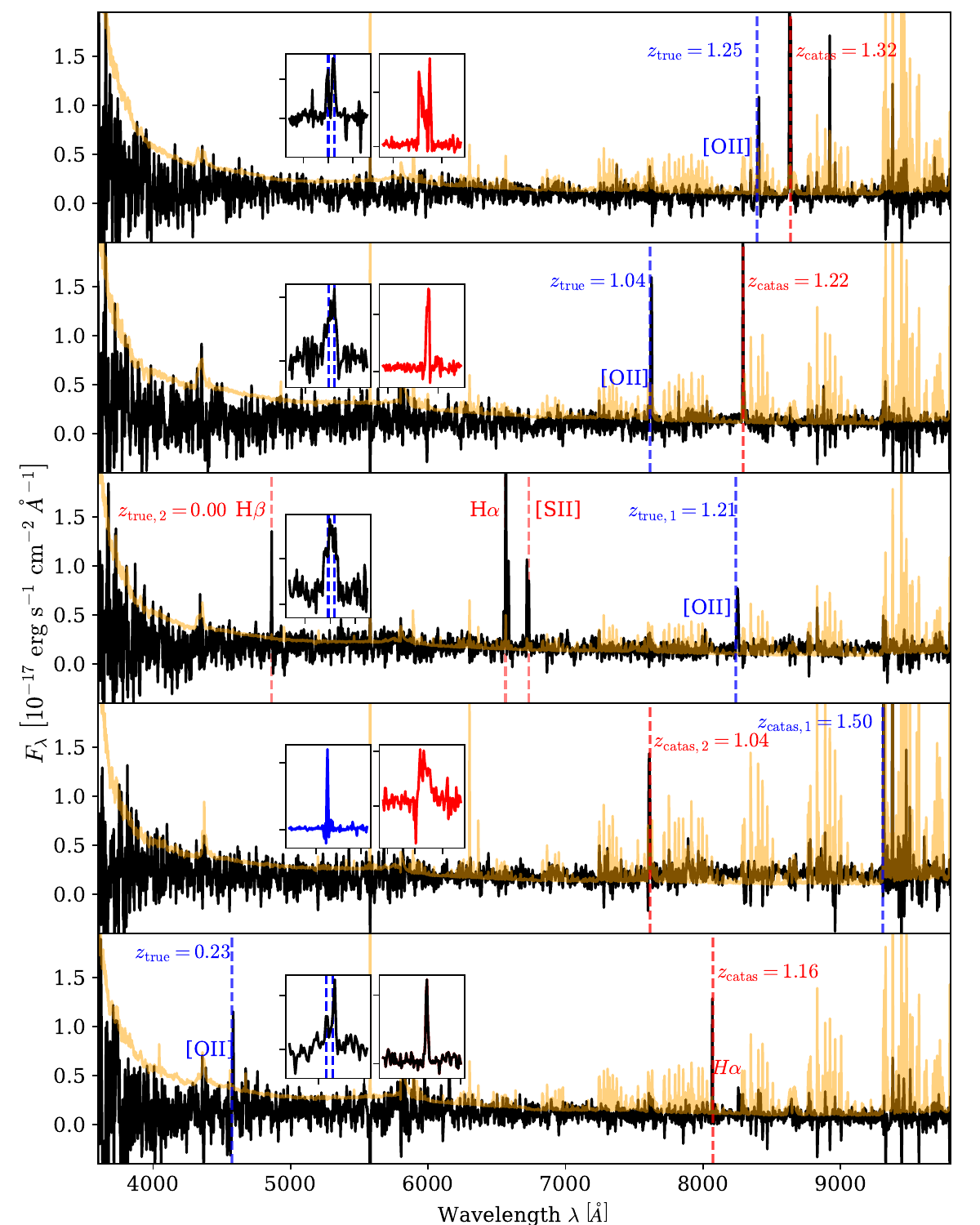}
\caption{Five ELG spectra that suffer from catastrophics in redshift measurements due to the sky confusion at $z_{\rm catas}\approx1.32$, other sky confusion patterns, double objects in the spectrum, total failures in the redshift measurement, and the other failures like the misidentification of H$\,\alpha$ to be \oii. The calibrated and sky-subtracted spectra are in solid lines. The orange shades in the foreground indicate the sky spectra. The blue vertical lines with $z_{\rm true}$ and \oii labels indicate the true \oii emissions positions except for the fourth spectra, in which the blue line indicates the sky residuals misidentified as \oii. The zoom-in spectra in their vicinity are presented in the inserted subplot on the left-hand side. The red vertical lines with $z_{\rm catas}$ point out the position of features misidentified as \oii line, except for the third spectra. Their enlarged features are the right-hand side inserted subplots. In the third spectra, there are H$\,\beta$, H$\,\alpha$, [S\,\textsc{ii}] to determine the redshift of the second object to be $z_{\rm true,2}=0$.  }
\label{fig:catas_vi}
\end{figure}

We present the ELG spectra of sky confusion at $z_{\rm catas}\approx 1.32$, sky confusion at other redshifts, double objects in the spectra, totally failed redshift measurements, and the misidentification of H$\,\alpha$ to be an \oii doublet in \cref{fig:catas_vi} as introduced in \cref{sec:catas_feature}. The insert subplots in the first, second and fifth spectra are enlarged spectral lines. The subplot on the left shows the spectra around the true \oii doublet (black solid lines with blue dashed vertical lines for \oii) and the one on the right is the false \oii line that results in catastrophics. The physical \oii emission lines for the left subplot in the first and the second spectra differ from the sky residuals in the right subplot. The H$\,\alpha$ emission in the right subplot of the fifth spectra was also different from the \oii doublet. However, \textsc{Redrock} was unable to distinguish them for some reason. Both redshift measurements are true in the third spectrum: one is determined by multiple emission lines from the $z_0=0$ object, and the other is from \oii emission. In the fourth spectra, both redshift measurements are wrong as they regard sky residuals at different redshifts as \oii doublets. All of them pass the good-redshift selection and most of them are within $0.8<z<1.6$ redshift range. This indicates that the 0.26\% is probably a lower limit of the catastrophics rate because there can be consistently wrong redshift measurements that pass the good-redshift criteria hidden in the current catalogue. %due to unidentified wrong redshifts from repeated observations.

\section{Catastrophics Impacts on Small-Scale 2PCF}
\label{append:small_scale}
\begin{figure}
\centering
\includegraphics[width=\textwidth]{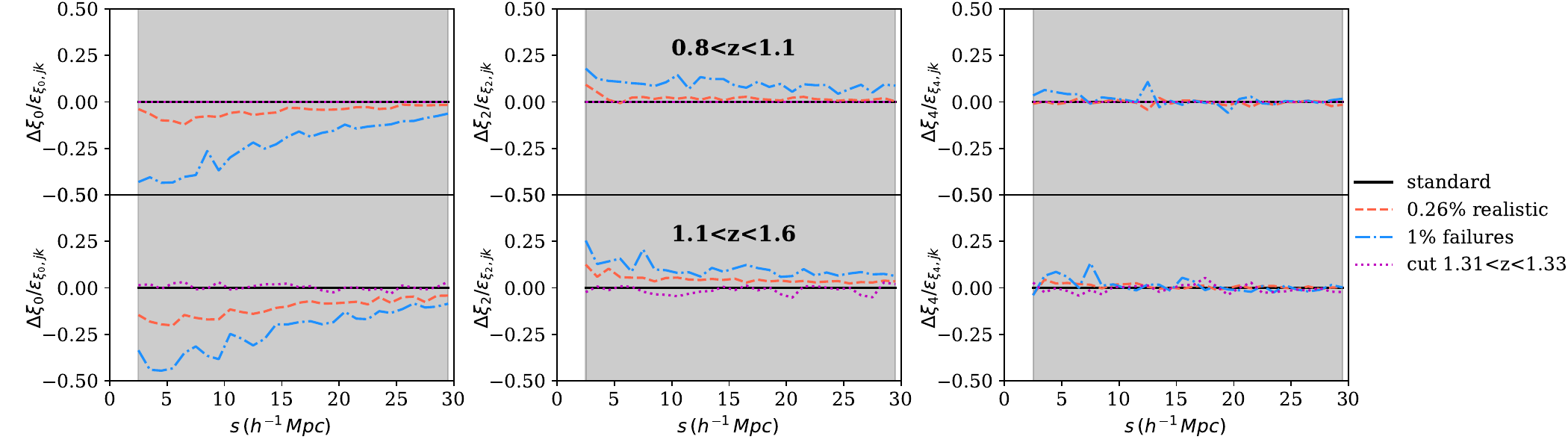}
\caption{Similar to \cref{fig:wzfail_effect_mps} but for the comparison between the standard \abacus galaxy mocks and the catastrophics-contaminated mocks with 0.26\% realistic catastrophics (dashed lines), 1\% random catastrophics (dash-dotted lines) and no ELGs at $1.31<z<1.33$ (dotted lines) at $s<30\mpch$. The grey regions are the analytical 1-$\sigma$ error of the ELG clustering. The impact of catastrophics at $1.1<z<1.6$ can be as large as $\sim0.5\varepsilon_\xi$, but is much smaller at $0.8<z<1.1$. }
\label{fig:catas_small}
\end{figure}

In \cref{sec:catas}, we demonstrate that the impact of catastrophics on cosmological scales in the configuration space is negligible. In this section, we discuss the influence of catastrophics on small scales as it has not been corrected in the current clustering. As the analytical covariance is no longer applicable at $s<20\mpch$, we used the jackknife error from the EDR ELG samples calculated with 128 subsamples using \textsc{pycorr} in our discussion. At $1.1<z<1.6$, the catastrophics can suppress the monopole by up to $\sim0.5\varepsilon_\xi$ and change quadrupole systematically as shown in \cref{fig:catas_small}. This indicates that the galaxy-halo connection studies of ELGs in galaxy surveys, with HOD or SubHalo Abundance Matching (SHAM), may need to consider this effect when they fit the observed ELG clustering at small scales.

\section{Author Affiliations}
\label{appendix:affiliations}

$^{1}${Institute of Physics, Laboratory of Astrophysics, \'Ecole Polytechnique F\'ed\'erale de Lausanne (EPFL), Observatoire de Sauverny, CH-1290 Versoix, Switzerland} 

\noindent \hangindent=.5cm $^{2}${Center for Cosmology and AstroParticle Physics, The Ohio State University, 191 West Woodruff Avenue, Columbus, OH 43210, USA}

\noindent \hangindent=.5cm $^{3}${Department of Astronomy, The Ohio State University, 4055 McPherson Laboratory, 140 W 18th Avenue, Columbus, OH 43210, USA}

\noindent \hangindent=.5cm $^{4}${The Ohio State University, Columbus, 43210 OH, USA}

\noindent \hangindent=.5cm $^{5}${Universit\'e Paris-Saclay, CEA, IRFU,  F-91191 Gif-sur-Yvette, France}

\noindent \hangindent=.5cm $^{6}${University of Michigan, Ann Arbor, MI 48109, USA}

\noindent \hangindent=.5cm $^{7}${Aix Marseille Universit\'e, CNRS, LAM (Laboratoire d'Astrophysique de Marseille) UMR 7326, F13388, Marseille, France}

\noindent \hangindent=.5cm $^{8}${Department of Physics and Astronomy, University of Waterloo, 200 University Ave W, Waterloo, ON N2L 3G1, Canada}

\noindent \hangindent=.5cm $^{9}${Perimeter Institute for Theoretical Physics, 31 Caroline St. North, Waterloo, ON N2L 2Y5, Canada}

\noindent \hangindent=.5cm $^{10}${Waterloo Centre for Astrophysics, University of Waterloo, 200 University Ave W, Waterloo, ON N2L 3G1, Canada}

\noindent \hangindent=.5cm $^{11}${Graduate Institute of Astrophysics and Department of Physics, National Taiwan University, No. 1, Sec. 4, Roosevelt Rd., Taipei 10617, Taiwan}

\noindent \hangindent=.5cm $^{12}${Center for Astrophysics $|$ Harvard \& Smithsonian, 60 Garden Street, Cambridge, MA 02138, USA}

\noindent \hangindent=.5cm $^{13}${Lawrence Berkeley National Laboratory, 1 Cyclotron Road, Berkeley, CA 94720, USA}

\noindent \hangindent=.5cm $^{14}${Physics Dept., Boston University, 590 Commonwealth Avenue, Boston, MA 02215, USA}

\noindent \hangindent=.5cm $^{15}${Department of Physics \& Astronomy, University College London, Gower Street, London, WC1E 6BT, UK}

\noindent \hangindent=.5cm $^{16}${Instituto de F\'{\i}sica, Universidad Nacional Aut\'{o}noma de M\'{e}xico,  Cd. de M\'{e}xico  C.P. 04510,  M\'{e}xico}

\noindent \hangindent=.5cm $^{17}${NSF NOIRLab, 950 N. Cherry Ave., Tucson, AZ 85719, USA}

\noindent \hangindent=.5cm $^{18}${Department of Physics \& Astronomy and Pittsburgh Particle Physics, Astrophysics, and Cosmology Center (PITT PACC), University of Pittsburgh, 3941 O'Hara Street, Pittsburgh, PA 15260, USA}

\noindent \hangindent=.5cm $^{19}${Kavli Institute for Particle Astrophysics and Cosmology, Stanford University, Menlo Park, CA 94305, USA}

\noindent \hangindent=.5cm $^{20}${SLAC National Accelerator Laboratory, Menlo Park, CA 94305, USA}

\noindent \hangindent=.5cm $^{21}${Departamento de F\'isica, Universidad de los Andes, Cra. 1 No. 18A-10, Edificio Ip, CP 111711, Bogot\'a, Colombia}

\noindent \hangindent=.5cm $^{22}${Observatorio Astron\'omico, Universidad de los Andes, Cra. 1 No. 18A-10, Edificio H, CP 111711 Bogot\'a, Colombia}

\noindent \hangindent=.5cm $^{23}${Institut d'Estudis Espacials de Catalunya (IEEC), 08034 Barcelona, Spain}

\noindent \hangindent=.5cm $^{24}${Institute of Cosmology and Gravitation, University of Portsmouth, Dennis Sciama Building, Portsmouth, PO1 3FX, UK}

\noindent \hangindent=.5cm $^{25}${Institute of Space Sciences, ICE-CSIC, Campus UAB, Carrer de Can Magrans s/n, 08913 Bellaterra, Barcelona, Spain}

\noindent \hangindent=.5cm $^{26}${Department of Physics, The Ohio State University, 191 West Woodruff Avenue, Columbus, OH 43210, USA}

\noindent \hangindent=.5cm $^{27}${School of Mathematics and Physics, University of Queensland, 4072, Australia}

\noindent \hangindent=.5cm $^{28}${Sorbonne Universit\'{e}, CNRS/IN2P3, Laboratoire de Physique Nucl\'{e}aire et de Hautes Energies (LPNHE), FR-75005 Paris, France}

\noindent \hangindent=.5cm $^{29}${Departament de F\'{i}sica, Serra H\'{u}nter, Universitat Aut\`{o}noma de Barcelona, 08193 Bellaterra (Barcelona), Spain}

\noindent \hangindent=.5cm $^{30}${Institut de F\'{i}sica d’Altes Energies (IFAE), The Barcelona Institute of Science and Technology, Campus UAB, 08193 Bellaterra Barcelona, Spain}

\noindent \hangindent=.5cm $^{31}${Instituci\'{o} Catalana de Recerca i Estudis Avan\c{c}ats, Passeig de Llu\'{\i}s Companys, 23, 08010 Barcelona, Spain}

\noindent \hangindent=.5cm $^{32}${Department of Physics and Astronomy, Siena College, 515 Loudon Road, Loudonville, NY 12211, USA}

\noindent \hangindent=.5cm $^{33}${Department of Physics and Astronomy, University of Sussex, Brighton BN1 9QH, U.K}

\noindent \hangindent=.5cm $^{34}${Department of Physics \& Astronomy, University  of Wyoming, 1000 E. University, Dept.~3905, Laramie, WY 82071, USA}

\noindent \hangindent=.5cm $^{35}${National Astronomical Observatories, Chinese Academy of Sciences, A20 Datun Rd., Chaoyang District, Beijing, 100012, P.R. China}

\noindent \hangindent=.5cm $^{36}${Departamento de F\'{i}sica, Universidad de Guanajuato - DCI, C.P. 37150, Leon, Guanajuato, M\'{e}xico}

\noindent \hangindent=.5cm $^{37}${Instituto Avanzado de Cosmolog\'{\i}a A.~C., San Marcos 11 - Atenas 202. Magdalena Contreras, 10720. Ciudad de M\'{e}xico, M\'{e}xico}

\noindent \hangindent=.5cm $^{38}${Space Sciences Laboratory, University of California, Berkeley, 7 Gauss Way, Berkeley, CA  94720, USA}

\noindent \hangindent=.5cm $^{39}${University of California, Berkeley, 110 Sproul Hall \#5800 Berkeley, CA 94720, USA}

\noindent \hangindent=.5cm $^{40}${Instituto de Astrof\'{i}sica de Andaluc\'{i}a (CSIC), Glorieta de la Astronom\'{i}a, s/n, E-18008 Granada, Spain}

\noindent \hangindent=.5cm $^{41}${Department of Physics, Kansas State University, 116 Cardwell Hall, Manhattan, KS 66506, USA}

\noindent \hangindent=.5cm $^{42}${Department of Physics and Astronomy, Sejong University, Seoul, 143-747, Korea}

\noindent \hangindent=.5cm $^{43}${CIEMAT, Avenida Complutense 40, E-28040 Madrid, Spain}

\noindent \hangindent=.5cm $^{44}${Space Telescope Science Institute, 3700 San Martin Drive, Baltimore, MD 21218, USA}

\noindent \hangindent=.5cm $^{45}${Department of Physics, University of Michigan, Ann Arbor, MI 48109, USA}

\noindent \hangindent=.5cm $^{46}${Department of Physics \& Astronomy, Ohio University, Athens, OH 45701, USA}

\noindent \hangindent=.5cm $^{47}${Department of Astronomy, Tsinghua University, 30 Shuangqing Road, Haidian District, Beijing, China, 100190}

\bibliographystyle{JHEP}
\bibliography{reference,DESI2024}



\end{document}